\documentclass[a4paper,fleqn]{cas-dc}

\usepackage[authoryear,longnamesfirst]{natbib}
\usepackage{bm}
\usepackage{colortbl}
\def\tsc#1{\csdef{#1}{\textsc{\lowercase{#1}}\xspace}}
\tsc{WGM}
\tsc{QE}

\begin{document}
\let\WriteBookmarks\relax
\def\floatpagepagefraction{1}
\def\textpagefraction{.001}

\shorttitle{GalPort}    

\shortauthors{V. Zozulia}  

\title[mode = title]{GalPort: Investigation of the bar in action-angle space}  

\tnotemark[1] 

\tnotetext[1]{} 

%

\author[1]{Viktor D. Zozulia}[                             orcid=0009-0007-5559-9777]
\cormark[1]
\fnmark[1]
\ead{vdzozulia.astro@gmail.com}

\credit{Conceptualization of this study, Methodology, Software}

\affiliation[1]{organization={St. Petersburg State University},
            addressline={Universitetskij pr. 28}, 
            city={St. Petersburg},
            postcode={198504}, 
            state={St. Petersburg},
            country={Russia}}


\begin{abstract}
\texttt{GalPort} is a Python package for analysing the orbital dynamics of evolving disc galaxy numerical models in action-angle space. The package implements novel numerical methods for \textcolor{black}{efficiently estimating actions}, angles, and frequencies across different, particle-specific timescales: on the scale of radial or vertical oscillation and on the resonant libration/circulation timescale. \textcolor{black}{The algorithm allows calculation of} these dynamic quantities simultaneously at all time steps of the simulated galactic evolution. With this tool, one can trace orbital behaviour within time-varying galactic potentials and classify orbits  (resonantly trapped, circulating, or passing through a resonance) based on their angle evolution. \texttt{GalPort} also includes specialised options for analysing the phase-space structure of a galactic bar in the disc plane and along its major axis. We demonstrate the package's performance on a typical $N$-body model of a barred galaxy, obtaining the global distributions of actions and frequencies and performing a detailed orbital decomposition. The code is publicly available under the MIT license at: \url{https://github.com/vdzozulia/galport}
\end{abstract}



\maketitle

\section{Introduction}
\label{sec: Intro}

Galaxies are complex, self-consistent dynamical systems in which stars and dark matter particles move along orbits in a gravitational potential that they generate themselves. This motion can lead to the evolution of the galactic density distribution and potential. In the dynamically cold disc galaxies, this can result in the formation of non-axisymmetric structures such as bars and spirals. The motion of stars within a galaxy can be described by various variables. To understand the dynamics completely, it is useful to adopt an approach that characterises the entire orbit. Many numerical simulations show that at least in a static potential, most of the orbits in such systems are periodic or quasiperiodic \citep{Binney_Spergel1982, Wang_etal2016, Machado_Manos2016, Smirnov_etal2021}, which means that they possess three integrals of motion \citep{Arnold1978}. Among the various functions of these integrals, the action variables $\bm{J}$ occupy a special place because, together with their conjugate angle variables $\bm{\theta}$, they form a complete set of canonical coordinates. The actions characterise the orbital shape, while the angles $\bm{\theta}$ increase linearly in time with a constant angular frequency $\bm{\Omega}$, tracking the position in the orbit. A detailed description of action–angle variables can be found in  \citet{Binney_Tremaine2008} (hereafter BT8).
\par
Over the past decade, action-angle variables have emerged as one of the most powerful tools for studying galactic dynamics. This has become possible due to the development of new algorithms that allow transformation between ordinary phase-space variables $(\bm{x}, \bm{v})$ and axisymmetric action-angle variables $(\bm{\theta}, \bm{J})$ \citep{Binney2012, Sanders_Binney2015, Binney_McMillan_2016}. At the same time, the emergence of user-friendly Python packages such as \texttt{AGAMA} and \texttt{Galpy} \citep{agama, galpy} has made these tools accessible to a wide range of researchers.
\par
The action-based approach is used to build equilibrium $N$-body models \citep{Sormani_etal2022, Hirashima_2025}. There are many studies focusing on the investigation of the solar neighbourhood using real and synthetic data from \textcolor{black}{Gaia} in action-angle space. In particular, they examine the general stellar distribution \citep{Trick_etal2016, Myeong_2018, Trick_etal2019, Binney_Vasiliev_2023, Binney_Vasiliev_2024}, and the motion of clusters and streams \citep{Fardal_2015, Vasiliev_2019, Reino_etal_2021, Arora_etal_2022, Malhan_etal2022, Sun_2023}. 
\par
Although there are methods for calculating action-angle variables in evolved non-axisymmetric systems, their consideration is limited to the dynamics of a single resonance \citep{Lichtenberg_Lieberman_1992, Binney_2018, Chiba_Schonrich_2022}. . Fortunately, even unperturbed axisymmetric action-angle variables can be used to investigate non-axisymmetric features of the potential by tracing their evolution in time, similar to how one monitors orbital parameters (semimajor axis, eccentricity, etc.)  in celestial mechanics. As a result, this approach has been successfully applied to study the properties of the bar and spirals in our Galaxy \citep{Binney_2020b, Trick_2022, Chiba_etal2021, Kawata2021, Chiba_etal2021b, Ghosh_etal2023, Dillamore_etal2024, Hunt_etal2019}. 
\par
Despite notable successes, only a relatively small number of papers have been devoted to studying the bar itself in the action-angle space  \citep{Wozniak2020, Debattista_etal2020, Zozulia_etal2024a, Zozulia_etal2024b, Zozulia_etal2025, Smirnov_etal2025}. However, most of the theoretical works on bar formation and evolution are written precisely using the language of action-angle variables \citep{ Quillen_etal2014, Polyachenko_Shukhman2020a, Chiba_Schonrich_2022, Chiba_2023}.  Basically, the bar is analysed using global parameters (such as the bar size and pattern speed $\Omega_\mathrm{p}$,  the amplitude of the second Fourier harmonic $A_2$ etc.) or by studying the motion of stars and gas in the ordinary phase space \citep{Fragkoudi_etal2025,  Ansar_2025}.  Alternatively, for the orbital study of bars, one applies frequency analysis  \citep{Parul_etal2020, Sellwood_Gerhard2020, Smirnov_etal2021, Tikhonenko_etal2021, Beraldo2023} or identifies the pure-resonant families of orbits \citep{Portail_etal2015a}.
\par
In this paper we begin with a brief introduction to action-angle variables in disc galaxies (Section~\ref{sec:act-ang}). In Section~\ref{sec:methods}, we present algorithms and the Python package \texttt{galport} for investigating the evolution of galactic models in the action-angle space. The main focus of these tools is on studying the structure and evolution of bars. The key features of the package are:

\begin{enumerate}
    \item The estimation of the action-angle variables and frequencies without short-term fluctuations on medium- and long-term timescales from time series of instantaneous action-angle or ordinary phase-space variables  (Sections~\ref{sec:actang_dift_theory} and \ref{sec:act_freq_dift}). In this case, all variables are calculated at once for all moments of the model's evolution from time series of ordinary coordinates and velocities. 
    \item Fast orbital classification based on the behaviour of the resonant angle combination (Sections~\ref{sec:orb_class} and \ref{sec:ang_decompose})
    \item Obtaining and analysing phase portraits in the action-angle space for galactic orbits in the plane of the bar and for near-bar orbits, as well as for orbits that are strongly aligned with the bar. (Sections~\ref{sec:pp_fit} and \ref{sec:PP_bar})
\end{enumerate}

 In previous papers \citet{Zozulia_etal2024a, Zozulia_etal2024b, Zozulia_etal2025}, we have already applied some of these tools to investigate the process of bar trapping, slow bar growth, and buckling. In this work, we provide additional examples of how to apply this package to the analysis of the $N$-body model in Section~\ref{sec:applications}.We present our conclusions in Section~\ref{sec:conclusion}, summarising the key results and outlining potential future developments of the \texttt{galport} package.

\section{Action-angle variables}
\label{sec:act-ang}


The motion of a star in a galactic potential  can be described by a Hamiltonian system. If this motion is regular, it can be described in terms of canonical action-angle variables. Actions $\bm{J}$ are integrals of motion and remain constant, while angles $\bm{\theta}$ increase linearly in time with an angular velocity $\bm{\Omega}$. Despite the convenience of action-angle variables, a general transformation between phase-space coordinates $(\bm{x},\bm{v})$ and action-angle variables $(\bm{\theta},\bm{J})$ is known only for a limited set of three-dimensional potentials.
\par
In galactic dynamics, spherical, cylindrical, and St{\"a}ckel potentials are among the most frequently considered. Their axial symmetry allows one to fix the $z$-component of the angular momentum. Most importantly, these potentials enable us to separate variables and find the third integral of motion. The process of calculating action-angle variables in such potentials is discussed in detail in BT08. Below, we briefly summarise the key results and then describe the behaviour of these variables in non-axisymmetric rotating potentials.

\subsection{Axisymmetric potentials}
\label{sec:act-ang-sym}
\textit{Spherical system.} Orbits in a spherical system potential $\Phi(r)$  conserve their angular momentum $\bm{L}$. Consequently, their orientation in space is maintained. The motion can be separated into angular and radial components within the orbital plane and is characterised by two actions: the total angular momentum $L = |\bm{L}|$ and the radial action. The radial action is found by integrating the radial momentum from pericentre  $r_p$ to apocentre $r_a$:  $J_r = 1/\pi \int_{r_p}^{r_a} p_r dr$. We also can use $L_z$ as the $z$-component of the angular momentum at the same coordinate system. Thus, action variables in spherical systems are given as $\bm{J} = (J_r, L - |L_z|, L_z)$. The calculation of the corresponding angle variables is described in BT08.
\par
  \textit{In a cylindrical system}, the $z$-component of the angular momentum $L_z$ and the energy of vertical motion are conserved. Therefore, we can introduce radial and vertical actions:

\begin{equation}
\label{eq:act_cyl}
    J_R = \dfrac{1}{2\pi} \oint v_R dR ;\;\; J_z = \dfrac{1}{2\pi}\oint v_z dz.
\end{equation}
In these equations, integration is carried out along the orbit between two successive apocentres or $z$ maxima. Following \citet{Lynden-Bell_Kalnajs1972, Binney_McMillan2011}, one can determine the radial, vertical and azimuthal frequencies:
\begin{equation}
\label{eq:freq_cyl}
    \textcolor{black}{\Omega_R} = \dfrac{2 \pi}{T_R};\;\; \textcolor{black}{\Omega_z} = \dfrac{2 \pi}{T_z};\;\; \textcolor{black}{\Omega_\varphi} = \dfrac{\Delta \varphi}{T_R} = \dfrac{1}{T_R} \int \dfrac{L_z}{R^2} dt
\end{equation}
where $T_R$ and $T_z$ are the time intervals between two successive
passages of apocenters and $z$-maxima, respectively. $\Delta \varphi$ is a change of azimuthal angle after one radial
oscillation. \textcolor{black}{Note that the angular speed $\Omega = L_z/R^2$ does not equal the azimuthal frequency.} In this case, the angles can be calculated as follows:
\begin{equation}
\label{eq:ang_cyl}
\theta_R = \textcolor{black}{\Omega_R} (t - t^R_{0});\;\; \theta_z = \textcolor{black}{\Omega_z} (t - t^z_0);\;\; \theta_\varphi = \Omega (t - t^R_0) + \varphi_0.
\end{equation}
Here $t^R_{0}$ and $t^z_0$ are the time moments of the first apocentre and first $z$ maximum passages, and $\varphi_0$ is the azimuthal angle of the first apocentre. \textcolor{black}{Note that, as in previous works \citep{Zozulia_etal2024a, Zozulia_etal2024b, Zozulia_etal2025}, for convenience we use a slightly non-standard notation.}
\par
\textit{The St{\"a}ckel system.} The St{\"a}ckel potential is the most general separable axisymmetric potential and has the following form:
\begin{equation}
\Phi(u, v) = \dfrac{U(u) - V(v)}{\sinh^2{u} + \sin^2{v}} \,, 
\end{equation}
where $u$ and $v$ are hyperbolic coordinates related to Cartesian coordinates as follows:
\begin{equation}
R = \Delta \sinh{u} \sin{v};\; z = \Delta \cosh{u} \cos{v}.
\end{equation}
In this potential, an orbit is confined to a region bounded by confocal ellipses on two opposite sides and confocal hyperbolas on the other two. The distance between the foci is $\Delta$. The expressions for actions and frequencies in this potential are described in BT8. 
\par
Numerous methods have been developed to estimate action-angle variables in realistic axisymmetric systems, as reviewed in~\citep{Sanders_Binney2015}. Among them, convergent and non-convergent ones are distinguished. Convergent methods present true action-angles as a series of the so-called toy action-angles, that are founded analytically in the toy potential (the isochrone one or harmonic oscillator) \citep{McGill_Binney_1990}. It allows for an accurate estimation of action angles over a relatively long period of time. At the same time, non-convergent methods do not have such accuracy, but they work much faster. \textcolor{black}{These methods are based on the local approximation of a general axisymmetric potential by a potential for which action-angle variables can be expressed explicitly. For example, the
}
St{\"a}ckel fudge method \textcolor{black}{\citep{Sanders_2012}} locally \textcolor{black}{estimates} $\Delta$ and \textcolor{black}{approximates the} potential  with a St{\"a}ckel one. It is one of the most accurate and fastest among non-convergent methods. Further, we will use the implementation of this method from the \texttt{AGAMA} software package ~\citep{agama}, if we can obtain the potential of the system. The cylindrical approach (Eq.~\ref{eq:act_cyl}-\ref{eq:ang_cyl}) will be used if we only have the trajectory of a particle $(\bm{x}, \bm{v})$.


\subsection{Non-axisymmetric potentials}
\label{sec:act-ang-nonsym}

In Section~\ref{sec:act-ang-sym}, we discussed how to estimate the action-angle variables in an axisymmetric galactic potential. However, real galaxies often have non-axisymmetric features such as spiral arms and bars, which need to be taken into account when calculating the action-angle variables. Furthermore,  these structures can change their shape and rotation rate. Here, we consider the case of a non-axisymmetric perturbation that rotates with a constant angular speed $\Omega_\mathrm{p}$. This system can be described by a Hamiltonian in the following form:

\begin{equation}
\label{eq:H_gal}
    H(\bm{v}, R, z, \phi) = \dfrac{1}{2}\bm{v}^2 - \Omega_\mathrm{p} L_z + \Phi(R,z) + \delta \Phi(R,z,\phi), 
\end{equation}
where  $\Phi(R,z)$ and $\delta\Phi(R,z,\phi)$ are the axisymmetric and non-axisymmetric parts of a potential, respectively. \textcolor{black}{It is assumed that the non-axisymmetric component is a small perturbation, such that $\delta\Phi(R,z,\phi)\ll \Phi(R,z)$}. $\phi = \varphi - \Omega_\mathrm{p} t$ is an azimuthal angle in a rotating reference frame. In terms of unperturbed action-angle variables of an axisymmetric Hamiltonian (cylindrical or St{\"a}ckel system) $\bm{J}=(J_R, \,J_z, \,L_z)$ and $\bm{\theta}=(\theta_R, \,\theta_z,\, \theta_\phi= \theta_\varphi- \Omega_\mathrm{p} t)$ the Hamiltonian of the system takes the following form:
\begin{equation}
\label{eq:pert_H}
    H(\bm{J}, \bm{\theta}) = H_0(\bm{J}) + \epsilon H_1(\bm{J}, \bm{\theta}) = H_0(\bm{J}) + \sum_{\bm{n}} h_{\bm{n}}(\bm{J}) e^{\bm{n}\cdot\bm{\theta}}, 
\end{equation}
here $\bm{n} = (n_R, n_z, n_\phi)$ is a triple of integers. Following the Hamilton's equations \citep{Arnold1978}, we can obtain the equations for derivatives of actions and angles:
\begin{equation}
\label{eq:pert_J}
    \dot{\bm{J}}(\bm{J}, \bm{\theta}) = -i\sum_{\bm{n}} \bm{n}\cdot h_{\bm{n}}(\bm{J}) e^{i\bm{n}\cdot\bm{\theta}};
\end{equation}
\begin{equation}
\label{eq:pert_theta}
    \dot{\bm{\theta}}(\bm{J}, \bm{\theta}) = \bm{\Omega}(\bm{J}) + \sum_{\bm{n}} \dfrac{\partial h_{\bm{n}}(\bm{J})}{\partial \bm{J}} e^{i\bm{n}\cdot\bm{\theta}}.
\end{equation}

\textcolor{black}{Here, the sums on the right-hand sides of these equations are of the first order of smallness. If the orbit is located far from resonances, one can solve this system to first-order accuracy. For this purpose, new invariant action-angle variables $(\bm{J}',\bm{\theta}')$ can be introduced.  Within this approximation, they satisfy $\bm{J}' = \mathrm{const}$ and $\bm{\theta}' = \bm{\Omega}(\bm{J}')t+\bm{\theta}_0$. Taking into account the expansion of the frequencies:
\begin{equation}
\label{eq: pert_omega}
\bm{\Omega}(\bm{J}' + \Delta \bm{J}') = \bm{\Omega}(\bm{J}') + \dfrac{ \partial \bm{\Omega}(\bm{J}')}{\partial \bm{J}} \Delta\bm{J}',
\end{equation}
and substituting Eq.~\eqref{eq: pert_omega} into Eq.~\eqref{eq:pert_theta} while replacing $\bm{J}$ with $\bm{J}'$ in the perturbation terms of Eqs.~\eqref{eq:pert_J} and \eqref{eq:pert_theta}, and subsequently integrating both sides with respect to time, one can derive the canonical transformation between the old axisymmetric variables $(\bm{J}, \bm{\theta})$ and the new invariant variables. To first-order accuracy, this transformation yields:}
\begin{equation}
\label{eq:nonres_J}
    \bm{J} = \bm{J}' - \sum_{\bm{n}}{\dfrac{\bm{n}\cdot h_{\bm{n}}(\bm{J}')}{\bm{n}\cdot\bm{\Omega}(\bm{J}')} e^{i\bm{n}\cdot \bm{\theta}'}};
\end{equation}
\begin{equation}
\label{eq:nonres_theta}
    \bm{\theta} = \bm{\theta}' - i \sum_{\bm{n}}{\dfrac{\partial h_{\bm{n}}(\bm{J'})}{\partial \bm{J}} \dfrac{e^{i\bm{n}\cdot \bm{\theta}'}}{\bm{n}\cdot\bm{\Omega}(\bm{J}')}} + i\dfrac{ \partial \bm{\Omega}(\bm{J}')}{\partial \bm{J}} \cdot \sum_{\bm{n}}{\dfrac{\bm{n}\cdot h_{\bm{n}}(\bm{J}')}{\left(\bm{n}\cdot\bm{\Omega}(\bm{J}') \right)^2} e^{i\bm{n}\cdot \bm{\theta}'}}.
\end{equation}

\textcolor{black}{We emphasise that the assumption $\bm{J}' = \mathrm{const}$ is a first-order perturbation theory approximation and holds only locally in the phase space, specifically for regular orbits far from major resonant and chaotic regions where the invariant tori are preserved.}
As we can see, the expressions~\eqref{eq:nonres_J}--\eqref{eq:nonres_theta}
tend to infinity at resonance, where $\textcolor{black}{\bm{n}}\cdot\bm{\Omega}(\bm{J}') = 0$, and cannot be used in its vicinity. This is a well-known problem of small denominators. To solve this issue, new \textcolor{black}{resonance} angle variables $\bm{\theta}'' = (\theta_{\mathrm{fast},1}, \,\theta_{\mathrm{fast},2},\, \textcolor{black}{\theta_{\mathrm{res},3}})=(\theta_R, \theta_z, \bm{N}\cdot \bm{\theta})$ are introduced near the specific resonance. \textcolor{black}{Here, $\bm{N}$ is a triplet of integers defining the resonance condition} $\bm{N}\cdot\bm{\Omega}(\bm{J}'_{\mathrm{res}}) = 0$.  Following the generated function
\begin{equation}
    S(\bm{J}'', \bm{\theta}) = \theta_R J_1 + \theta_z J_2 +
     \bm{N}\cdot \bm{\theta} \,J_3
\end{equation}
new action $\bm{J}'' = (J_1,\, J_2,\,J_3)$ are obtained as 
\begin{equation}
\bm{J}'' =\dfrac{\partial S}{\partial \bm{\theta}} = \left( J_R - \dfrac{n_R}{n_\phi} L_z,\,J_z - \dfrac{n_z}{n_\phi} L_z,\, \dfrac{L_z}{n_\phi} \right)
\end{equation}
\par
If we average the Hamiltonian (\ref{eq:pert_H}) by fast angles $\theta_{\mathrm{fast},1}$ and $\theta_{\mathrm{fast},2}$, it takes the following form:
\begin{equation}
\label{eq:H_aver}
    \overline{H}(\bm{J}'', \textcolor{black}{\theta_{\mathrm{res},3}})  = H_0(\bm{J}'') +  \sum_{k} h_{k\bm{N}} (\bm{J}'') e^{ik \textcolor{black}{\theta_{\mathrm{res},3}}}
\end{equation}
As follows from the \textcolor{black}{Hamilton}-Jacobi equation,
$J_1$ and $J_2$ are new integrals of motion and $J_3$ can be found from Equation~(\ref{eq:H_aver}). 
If we keep only the first term of the sum and expand all functions into a Taylor series to the second order near the fixed equilibrium point, where \textcolor{black}{$\dot{\theta}_{\mathrm{res},3}(\bm{J}''_\mathrm{eq}, \theta_{\mathrm{eq,3}}) =0$ and $\dot{J}_{3}(\bm{J}''_\mathrm{eq}, \theta_{\mathrm{eq,3}}) =0$}, we obtain \textcolor{black}{the following}:
\begin{equation}
\label{eq:H_pend}
    \overline{H}(\bm{J}'', \theta_3)\approx \dfrac{1}{2} G (\bm{J}''_{\mathrm{eq}})(J_3 - J_{\mathrm{eq},3})^2 + 2 h_{\bm{N}} (\bm{J}''_{res}) \cos{(\theta_3 - \textcolor{black}{\theta_{\mathrm{eq},3}})},
\end{equation}
where $G(\bm{J}\textcolor{black}{''}) = \partial^2{H_0}/\partial{J_3}^2 = \partial{\Omega_3}/\partial{J_3}$. \textcolor{black}{Up to numerical coefficients, this Hamiltonian coincides with the Hamiltonian of a nonlinear pendulum.}  \textcolor{black}{It is known that for this problem the resonant angle exhibits three types of behaviour \citep{Lichtenberg_Lieberman_1992}: circulation with positive or negative angular velocities and libration about a fixed point, where the angular velocity is near zero.} On the one hand, we can solve this one-dimensional system numerically to obtain the time evolution of $J_3(t)$ and $\textcolor{black}{\theta_{\mathrm{res},3}}(t)$. On the other hand, we can \textcolor{black}{approach} it analytically by introducing new action-angle variables for the pendulum $(\bm{J}^p, \bm{\theta}^p)$ \citep{Lichtenberg_Lieberman_1992, Binney_2018, Chiba_Schonrich_2022}.  
\par
Thus, we can outline a procedure to identify the axisymmetric action-angle variables that correspond to a resonant or near-resonant torus. First, we introduce and fix the pendulum actions and angles $(\bm{J}^p, \bm{\theta}^p)$ for the Hamiltonian (\ref{eq:H_pend}) or find the solution of the more general system (\ref{eq:H_aver}) using direct integration. In this way, we fix $\overline{H}$, $J_1$, $J_2$ and numerically calculate all possible values of $J_3$ and $\textcolor{black}{\theta_{\mathrm{res},3}}$. The angle variables $\theta_{\mathrm{fast},1}$ and $\theta_{\mathrm{fast},2}$ can take any value from $0$ to $2\pi$. Then, we transform to the non-resonant action-angle variables, accounting for the non-resonant terms ($\bm{n} \not\in {k\bm{N}: k\in \mathbb{Z}}$) of the perturbation given by Eqs.~\eqref{eq:nonres_J} and~\eqref{eq:nonres_theta}. It is worth noting that this procedure can also work in the opposite direction: the pendulum's action-angle variables can be found from the ordinary coordinates and velocities.
\par
A significant difficulty in this approach lies in determining the terms $h_{\bm{n}}(\bm{J})$ and their partial derivatives. These terms can be obtained from a Fourier expansion of $H$ in Eq.~(\ref{eq:pert_H}). For this purpose, we fix the actions and vary the angles, then obtain the corresponding ordinary phase-space coordinates and calculate the Hamiltonian given by Equation~(\ref{eq:H_gal}). This approach is presented in \citet{Binney_2018} for analysing resonant phenomena in two-dimensional galactic potential\textcolor{black}{s} with a bar.  As proposed by \citet{Quillen_etal2014} and \citet{Zozulia_etal2024b} two resonances  play a  crucial role in the bar formation and evolution: \textcolor{black}{the inner Lindblad resonance (ILR, $\textcolor{black}{\Omega_\varphi} - \textcolor{black}{\Omega_R}/2 = \Omega_\mathrm{p}$), and the vertical inner Lindblad resonance (vILR, $\textcolor{black}{\Omega_\varphi} - \textcolor{black}{\Omega_z}/2 = \Omega_\mathrm{p}$)}. Unfortunately, the approach described above is not applicable in a system  with two resonances. Below, we consider how we can examine the three-dimensional bar in this way.

\subsection{Hamiltonian of a bar}
\label{sec:H_bar}
\textcolor{black}{In the disc plane, most of the bar-supporting orbits are trapped at the ILR. Their backbone is formed by the so-called $x_1$ family, which consists of periodic orbits elongated parallel to the bar's major axis. The main resonance playing a central role in the vertical dynamics of the bar structure is the vILR}  \citep{Quillen2002, Quillen_etal2014, Sellwood_Gerhard2020, Zozulia_etal2024b}. These resonances are characterised by the angles $2\theta_\phi - \theta_R$ and $\theta_z - 2\theta_\phi$ (or $\theta_z - \theta_R$; these two angles are approximately equal in a bar), respectively. We consider the bar Hamiltonian in the form of Eq.~(\ref{eq:pert_H}) and average it over only one fast variable ($\theta_R$ or $\theta_z$). The resonant angle combinations $\textcolor{black}{\theta_{\mathrm{res},1}} = \theta_z - \theta_R$ and $\textcolor{black}{\theta_{\mathrm{res},2}} = 2\theta_\phi - \theta_R$ evolve slowly for bar particles; in mechanical terms, they are slow variables, \textcolor{black}{while the angle variable $\theta_{R}=\theta_{\mathrm{fast},3}$ is the fast one.} Therefore, we can perform a canonical transformation, leading to the new action-angle variables
\begin{equation}
    \bm{J}^b = (J_z, L_z/2, J_v),\; \bm{\theta}^b = ( \textcolor{black}{\theta_{\mathrm{res},1}}, \textcolor{black}{\theta_{\mathrm{res},2}}, \theta_{\mathrm{fast},3}),
\end{equation}
here, $J_v=J_R + J_z + L_z/2$ is a new invariant (for more details, see Appendix B of \citet{Zozulia_etal2024a}). The averaged bar Hamiltonian can then be written as:

\textcolor{black}{
\begin{equation}
\label{eq:H_bar}
    H_{\mathrm{bar}}(\bm{J}^b, \bm{\theta}_\mathrm{res}) = \\ H_0(\bm{J}^b) + \sum_{\bm{m}} h_{\bm{m}} (\bm{J}^b) e^{i\bm{m}\cdot \bm{\theta}_\mathrm{res}},
\end{equation}
}

\textcolor{black}{here $\bm{m} = (m_1,\, m_2)$ is a pair of integers, and $\bm{\theta}_\mathrm{res} = (\textcolor{black}{\theta_{\mathrm{res},1}},\, \textcolor{black}{\theta_{\mathrm{res},2}})$.}
We have obtained the system of two actions and two angles, with a fixed $J_v = \mathrm{const}$. Let us now consider the physical significance of the terms in Eq.~({\ref{eq:H_bar}}). The term $H_0(\bm{J})$ corresponds to the axisymmetric Hamiltonian in the absence of resonances. Terms with $m_1=1$ and $m_1=3$ are responsible for the vertical asymmetry. \textcolor{black}{This asymmetry is related to the curvature of the Laplace plane, where vertical gravitational acceleration vanishes, and to} a bar or disc \textit{buckling} (the rapid process of losing vertical symmetry). The vertical resonance and banana-shaped orbits in the bar arise due to the influence of terms with  $m_1=2$. It should be noted that if $m_2 = 0$, the emerging perturbations appear in an axisymmetric disc. Terms $(m_1=0, m_2\neq 0)$ relate to the ILR or the bar. 
In Table \ref{tab:ang_Hbar}, we list the main terms of the Hamiltonian (\ref{eq:H_bar}) up to the seventh order in the non-resonant angles. Higher-order terms also influence the bar, but their effect is weaker. We also explicitly exclude from consideration terms associated with other resonances, such as $6\theta_z - 5\theta_R$, $4\theta_z - 3\theta_R$, $3\theta_z - 2\theta_R$, which correspond to pretzel-shaped and other resonant orbit families. We believe that these high-order resonances do not play a crucial role in the vertical evolution of the bar. They correspond to orbits with high $J_z$ value and occupy a smaller volume in action-angle space relative to vILR, at least during the early stages of the bar's evolution. However, further \textcolor{black}{research is} required for a more accurate bar description.
\par
The Hamiltonian introduced above describes nearly all the inner bar kinematics, including the width and location of resonances. By determining how this Hamiltonian changes and how it is affected by initial conditions and other resonances (mainly corotation and outer \textcolor{black}{Lindblad} resonance, OLR), we can understand the essence of the bar evolution. 
\par
Moreover, such Hamiltonian systems are simpler to explore than systems with three degrees of freedom. For example, one can plot Poincaré maps \textcolor{black}{--- surfaces of section that reduce the continuous dynamical system to a lower-dimensional discrete representation ---} to investigate the phase-space structure and chaotic orbital behaviour arising from resonance overlap.

Moreover, such Hamiltonian systems are simpler to explore than systems with three variables. For example, one can plot Poincaré maps to investigate the phase-space structure and chaotic orbital behaviour arising from resonance overlap.  In practice, however, determining the functions $h_{\bm{m}} (\bm{J}^b)$ is challenging because it requires a computationally expensive three-dimensional Fourier transform on a three-dimensional grid in action space. Therefore, we propose an alternative, simpler approach for exploring the phase space of the bar.
\par

\begin{table}[h!]
\caption{Main terms of the bar Hamiltonia up to 7-th order at non-resonant angles $\theta = m_1(\theta_z-\theta_R) + m_2(2\theta_\phi-\theta_R)$. }                 
\label{tab:ang_Hbar}    
\centering                        
\begin{tabular}{c c c c}      
\hline\hline               
$(m_1, m_2)$ & Angle & Order & Description \\         
\hline                      
   $(0,0) $& $-$ & $0$ & axisymmetric $H_0$ \\    
   $(1,0)$ & $\theta_z - \theta_R$ & $2$ & axisymmetric  buckling  \\
   $(2,0)$ & $2\theta_z - 2\theta_R$ & $4$ & {axisymmetric vILR} \\
   $(3,0)$ & $3\theta_z - 3\theta_R$ & $6$ & vertical asymmetry   \\
\hline
   $(0,1)$ & $2\theta_\phi - \theta_R$ & $3$ & ILR \\
   $(0,2)$ & $4\theta_\phi - 2\theta_R$ & $6$ & ILR \\
   
\hline
   $(1,-1)$ & $\theta_z - 2\theta_\phi$ & $3$ & buckling \\
   $(1,1)$ & $\theta_z + 2\theta_\phi - 2\theta_R$ & $5$ & buckling \\
   $(1,-2)$ & $\theta_z - 4\theta_\phi+\theta_R$ & $6$ & buckling \\
   $(3,-1)$ & $3\theta_z - 2\theta_\phi+2\theta_R$ & $7$ &  vertical asymmetry\\
   
\hline   
   $(2,-2)$ & $2\theta_z - 4\theta_\phi$ & $6$ & vILR \\
   $(2,-1)$ & $2\theta_z - 2\theta_\phi - \theta_R$ & $5$ & vILR \\
   $(2,1)$ & $2\theta_z + 2\theta_\phi - 3\theta_R$ & $7$ & vILR \\

\hline                                  
\end{tabular}
\end{table}
Let us consider only orbits \textcolor{black}{whose apocentres lie in the plane that contains the $z$-axis and the major axis of the bar (their projection on the $xy$-plane aligns along the bar), hereafter bar-aligned 3D orbits}. This allows us to reduce the system~(\ref{eq:H_bar}) to the system with one action and one angle. We have two additional equations: $\dot{\theta}_{\textcolor{black}{\mathrm{res},2}}(\bm{J}^b, \bm{\theta}^b) = 0$ and $\textcolor{black}{\theta_{\mathrm{res},2}} = 0$ for this system.  \textcolor{black}{This means that these orbits are exactly at the ILR without libration.}  Therefore, we can incorporate these constraints, fix the Jacobian integral $H$ (we also can fix $J_v$), and by considering the averaged Hamiltonian dynamics (which excludes fast variables $\theta_R$ or $\theta_z$), we can reduce the six-dimensional system to a two-dimensional system. Let these two variables be $J_z$ and $\textcolor{black}{\theta_{\mathrm{res},1}}=\theta_z-\theta_R$. We can then write the general Hamiltonian for this system, up to order $n_{\mathrm{max}}$ in the angle $\textcolor{black}{\theta_{\mathrm{res},1}}$ as: 
\begin{equation}
\begin{split}
\label{eq:H_2d}
    H_{\mathrm{2d}}(J_z, &\textcolor{black}{\theta_{\mathrm{res},1}}) = h_0(J_z) + \\ &\sum_{n=1}^{n_\mathrm{max}}{h_{s,n} (J_z) \sin{(n\textcolor{black}{\theta_{\mathrm{res},1}})} + h_{c,n} (J_z) \cos{(n\textcolor{black}{\theta_{\mathrm{res},1}})}}.
\end{split}
\end{equation}

This Hamiltonian has a simple form and allows us to explore the vertical structure of the bar in two-dimensional phase space $(J_z, \theta_z-\theta_R)$. In subsequent sections, we will provide the algorithm for calculating average action-angle variables and determining the set of functions $h_0(J_z)$, $h_{c,n}(J_z)$, $h_{s,n}(J_z)$.

\section{Numerical methods}
\label{sec:methods}

\subsection{Action-angle calculation at different time-scales}
\label{sec:actang_dift_theory}

As shown in Section \ref{sec:act-ang-nonsym}, the action-angle variables in a non-axisymmetric potential change due to the influence of perturbative terms in the Hamiltonian of the system. This change occurs on different time scales, with fast variations related to fast angles (for example, $\theta_R$ and $\theta_z$), and mean-term variations associated with averaged movement near resonances (such as ILR, vILR, corotation, and etc.). In general, as a result of secular evolution of galaxies, the Hamiltonian of the system also evolves, and these resonances move across the phase space.
In this section we show how to estimate actions, angles and frequencies at these different time scales (fast-term or instantaneous, mean-term or averaged and secular ones).

\subsubsection{Instantaneous variables}
\label{sec:inst_act}

We estimate the instantaneous actions in an axisymmetric potential using the St{\"a}ckel fudge method (\texttt{agama.ActionFinder}) from the \texttt{AGAMA} package \citep{agama}. To obtain the axisymmetric potential of an $N$-body model, the potential is first expanded into a series of basis functions, from which the axisymmetric part is then extracted. This expansion can be done using either spherical or cylindrical functions, depending on the specific goals of the analysis. The cylindrical expansion more accurately describes the potential of a galactic disc; however, it involves higher computational costs. In contrast, the multipole expansion is less accurate but significantly faster.  Therefore, for detailed analysis of a limited number of potentials, we employ cylindrical basis functions, whereas to study the evolution of action variables requiring numerous expansions of potentials, we use the multipole expansion. All these procedures are also produced by the \texttt{AGAMA} package \citep{agama}.
\par
Thus, this method calculates three actions, three frequencies, and three angles. Despite its high computational efficiency and broad applicability to galactic models \citep{Debattista_etal2020, Debattista_etal2025}, it has several limitations when applied to non-axisymmetric systems.
\par
The first limitation concerns the use of axisymmetric frequencies. We caution against employing these frequencies to examine non-axisymmetric systems, as they do not accurately represent the true time derivatives of the angles, even when averaged over time. This inaccuracy arises because they are derived from only the axisymmetric part of the Hamiltonian (\ref{eq:pert_H}). Therefore, we consider expression (\ref{eq:pert_theta}) and an orbit close to resonance $\textcolor{black}{\theta_{\mathrm{eq},3}} = \bm{N} \cdot \bm{\theta} = const$. The averaged time angle derivative is given as follows:
\begin{equation}
   \left< \dot{ \bm{\theta}} \right> = \left< \bm{\Omega}(\bm{J}(t)) \right> + \sum_k \left< \dfrac{\partial h_{k\bm{N}}(\bm{J}(t))}{\partial \bm{J}} \cos{k(\textcolor{black}{\theta_{\mathrm{res},3}}(t) - \textcolor{black}{\theta_{\mathrm{eq},3}})} \right>.
\end{equation}

\textcolor{black}{Here, the averaging is performed over timescales significantly longer than the variation timescales of the fast angles.} Note that \textcolor{black}{under this procedure,} the terms with fast angles are excluded, whereas the resonant ones are left. $\left< \bm{\Omega}(\bm{J}(t)) \right>$ is a set of averaged axisymmetric frequencies and differs from the true average values  $\left< \dot{ \bm{\theta}} \right>$ by the constant.  Therefore, we do not recommend using instantaneous frequencies, even for rough estimates of orbital frequencies. Near resonance, these estimates are likely to be systematically offset by a constant value that varies between individual orbits. Instead, the instantaneous frequencies can be approximated, for example, by the differences in instantaneous angles:

\begin{equation}
    \bm{\Omega}_{in}(t) = \dfrac{\bm{\theta}_{in}(t + \Delta t) - \bm{\theta}_{in}(t - \Delta t)}{2\Delta t}.
\end{equation}

To calculate this value, the orbit must be integrated over a short time interval in the rotating potential, advancing both forward and backward in time. If the integration interval is extended significantly beyond the libration or circulation period around the resonance, and the rotation of the angle  (transition through $2\pi$), then the orbital frequencies $\textcolor{black}{\Omega_R}$, $\textcolor{black}{\Omega_z}$ and $\textcolor{black}{\Omega_\varphi}$ can be estimated. This procedure reminds \textcolor{black}{of} the algorithm described in \cite{Ceverino_Klypin2007} for the angular frequency. The authors considered the evolution of the azimuthal angle and used the least squares method to approximate its evolution, thus finding its average rate of change. A similar procedure can be applied to the instantaneous angles $\theta_R$, $\theta_z$ and $\theta_\phi$. This method looks promising, but we will use a different procedure, which we will describe below in Section~\ref{sec:aver_act-ang}.
\par
One more limitation of using instantaneous values relates to their oscillations on short timescales $\Delta t \approx T_R$ or $T_z$. These fluctuations are caused by the influence of non-resonant terms in the Hamiltonian~(\ref{eq:pert_H}), which are connected to fast angle combinations. Specifically, it means that the instantaneous action for bar particles in the inner Lindblad resonance (ILR) does not preserve the adiabatic invariant, $J_f = J_R + L_z/2$. The action combination will oscillate around the mean value with a large amplitude, making it difficult to analyse the galaxy model in the action-angle space, blurring the basic structures. To eliminate these short-time fluctuations, we introduce averaged action-angle variables.
\par
It should be noted that counter-rotating stars ($L_z < 0$) must be considered separately. This is due to the fact that the precession rate for non-counter-rotating orbits is $\textcolor{black}{\Omega_\varphi} - \textcolor{black}{\Omega_R}/2$, while for counter-rotating ones it is equal to $\Omega_\varphi + \textcolor{black}{\Omega_R}/2$. Therefore, for orbits with $\textcolor{black}{\Omega_\varphi} < 0$, we replace the angular velocity $\textcolor{black}{\Omega_\varphi}$ by $\textcolor{black}{\Omega_\varphi} + \textcolor{black}{\Omega_R}$ and the angle $\theta_\varphi$ by $\theta_\varphi + \theta_R$.  In addition to angle, we need to perform the canonical change of the radial action $J_R$ to $J_R - L_z$. Next, we will assume these replacements for counter-rotating particles, by default.

\subsubsection{Averaged variables}
\label{sec:aver_act-ang}
\begin{figure*}
    \centering
    \includegraphics[width=1.0\linewidth]{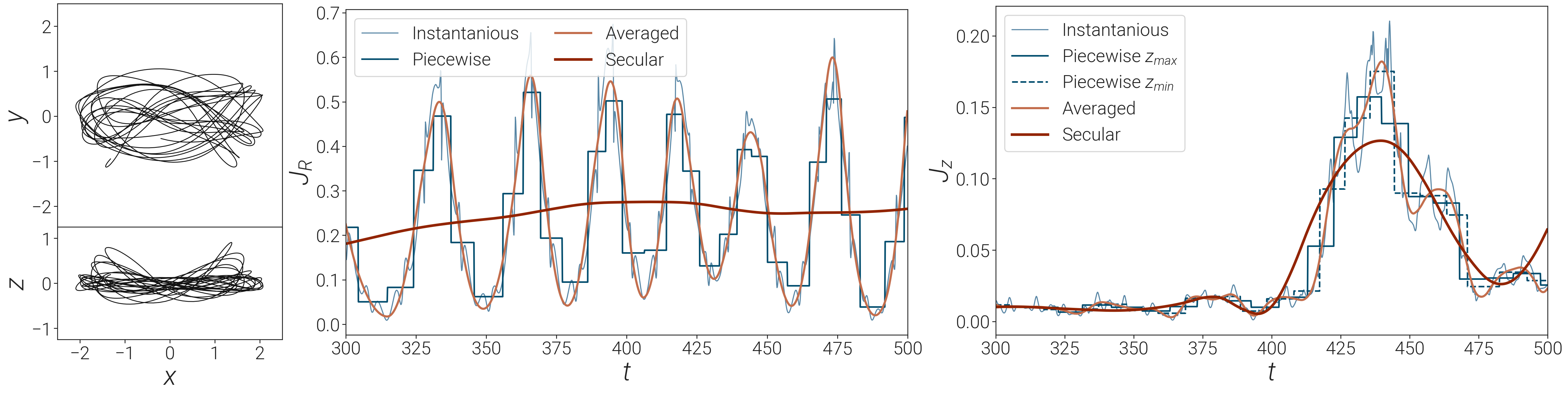}
    \caption{\textit{Left panel}: the orbit of the bar particle of the $N$-body model over the time interval from $t=300$ to $t=500$. \textit{The middle and right panels} show the evolution of $J_R$ and $J_z$ of this orbit on different time-scales (instantaneous or short-term, average or medium-term and secular or long-term). Piecewise function demonstrate the mean value of instantaneous action between apocentre for $J_R$ and between minima and maxima of $z$ for $J_z$.}
    \label{fig:aver_actions}
\end{figure*}

The next step in addressing the issue described above is to eliminate short-term oscillations. Various methods exist for smoothing signals and filtering out fluctuations, including B-splines, wavelet smoothing, moving average, and kernel-based smoothing and regression techniques such as LOESS \citep{Cleveland1979Loess}, the Savitzky–Golay method \citep{SavitzkyGolay1964}, and the Hodrick–Prescott decomposition \citep{HodrickPrescott1997}. Most of these regression methods require tuning external parameters, the most critical of which is the width of the smoothing window or kernel. However, our objective is to calculate action-angle variables free from short-term oscillations, effectively estimating resonance variables as given by the averaged Hamiltonian system (\ref{eq:H_aver}) or (\ref{eq:H_bar}). Since each orbit oscillates and librates at different frequencies, it is not feasible to select a single constant window or kernel width that is appropriate for all particles; each orbit demands an individualised approach.
\par
To this end, we developed a simple algorithm that averages instantaneous actions and frequencies individually for each orbit. This procedure has been employed in our previous works to study bar evolution \citep{Zozulia_etal2024a, Zozulia_etal2024b, Zozulia_etal2025}. Below, we describe the algorithm in detail. It consists of two straightforward steps: generation of piecewise functions of averaged variables and subsequent smoothing.
\par
 The first step is to obtain the mean value of the actions and frequencies between the required time intervals. The radial and angular actions $J_R$ and $L_z$ are averaged between two adjacent apocentres. We get the mean value $J_z$ between $z$-minima and $z$-maxima separately. Also, we find the mean value of the frequencies. Let us $t_{R, i}$, $t^{max}_{z,i}$, $t^{min}_{z,i}$ to be the passing time of $i$-th apocentre, $z$-maximum and minimum, respectively. Below, we provide the formulas to calculate the mean actions and frequencies between time moments $i$ and $i+1$:

\begin{align}
\begin{split}
\label{eq:mean_actfreq}
&\left< J_R \right>_i = \dfrac{1}{T^R_i} \int_{t_{R,i}}^{t_{R,i+1}} J_R dt;\\
&\left< L_z \right>_i = \dfrac{1}{T^R_i} \int_{t_{R,i}}^{t_{R,i+1}} L_z dt;\\
&\left< J_z^{max} \right>_i = \dfrac{1}{T_{z,i}^{max}} \int_{t^{max}_{z,i}}^{t^{max}_{z,i+1}} J_z dt; \\
&\left< J_z^{min} \right>_i =  \dfrac{1}{T_{z,i}^{min}} \int_{t^{min}_{z,i}}^{t^{min}_{z,i+1}} J_z dt; \\
&\left< \dot{\theta}_R \right>_i = \dfrac{2\pi}{T_{R,i}};\;\;\;\left< \dot{\theta}_\varphi \right>_i = \dfrac{1}{T_{R,i}} \int_{t_{R,i}}^{t_{R,i+1}} \dot{\varphi}dt = \dfrac{\Delta \varphi_i}{T_{R,i}}; \\
&\left< \dot{\theta}^{max}_z \right>_i = \dfrac{2\pi}{T_{z,i}^{max}};\;\;\; \left< \dot{\theta}^{min}_z \right>_i = \dfrac{2\pi}{T_{z,i}^{min}}.
\end{split}
\end{align}

 Here, the capital $T_i$ corresponds to the time between two specific time points $t_i$ and $t_{i+1}$; $\Delta\varphi$ is the polar angle covered by the particle between two neighbouring apocentres (if the $\Delta\varphi$ is negative, it is replaced by $\Delta\varphi + 2\pi$). 
 In fact, the algorithm yields a value for $\theta_R$ that is a multiple of $2\pi$ at apocentre. In $z$-coordinate, \textcolor{black}{we apply the algorithm to both the time interval between z-maxima and z-minima}
 to avoid preference to any one direction and remove fluctuations that are multiples of a half-period of oscillations along $z$, in particular, terms $h_{(0,\pm 1,0)}(\bm{J})$ in Eq.~(\ref{eq:pert_H}). In Appendix~\ref{ap:aver}, we explain how calculating the average value of a quantity can help eliminate perturbed terms that are proportional to $e^{n_1\theta_R}$ and $e^{n_2\theta_z}$ for corresponding variables and also significantly reduce the influence of fast terms of perturbed Hamiltonian. We also can use expressions (\ref{eq:act_cyl}) instead of (\ref{eq:mean_actfreq}) for estimation of the mean values of actions.
\par
The algorithm requires high time resolution to work correctly, but this is not always achieved in the $N$-body model. Therefore, we use a cubic spline to interpolate coordinates and instantaneous actions, thereby improving the accuracy of determining the time when a particle passes apocentres and reaches the maxima and minima of $z$.  In our models, the time resolution is $0.125$ \textcolor{black}{time units}. However, we increase it by a factor of $100$ using cubic splines. \textcolor{black}{For comparison, the characteristic period of bar rotation is approximately $30$ time units.} 
\par
The next step is to smooth the resulting piecewise-defined functions while maintaining their average values.  For this purpose, we use the mean-preserving quadratic spline well described in the paper \cite{Ruiz_Jose2022} and implemented by authors in the package {\tt{mpspline}}. In our work, we use it with minor modifications for our problem. Below, we briefly describe the algorithm.  
\par
Let us assume that we have known $n$ mean values $y_i$ for any function in the time interval $[t_{i},\,t_{i+1}]$ and we want to approximate the function in this interval using a quadratic polynomial:
\begin{equation}
\label{eq:spline0}
    S_i(t) = a_i t^2 + b_i t +c_i,\; t \in [t_{i},t_{i+1}] .
\end{equation}

For these purposes, we create a system of $3n$ linear equations for the coefficients $(a_i, b_i, c_i)$, which includes conditions of continuity, smoothness, mean preservation, and a boundary condition on the second derivative:

\begin{equation}
\label{eq:spline1}
    \begin{split}
    \int_{t_i}^{t_{i+1}} S_i(x) &= y_i \cdot (t_{i+1}-t_i),\; : \;i=1,2,...,n ;\\
    S_i(t_{i+1}) &= S_{i+1} (t_{i+1}) \; : \; i=1,2,...,n-1 ;\\
    S'_i(t_{i+1}) &= S'_{i+1} (t_{i+1}) \; : \; i=1,2,...,n-1 ;\\
    S''_1(t_{2}) &= S''_2(t_{2}) \; \mathrm{or} \; a_1 = a_2 ;\\
    S''_{n-1}(t_n) &= S''_n(t_n) \; \mathrm{or} \; a_{n-1} = a_n.
    \end{split}
\end{equation}

This system of linear equations can be solved using the tridiagonal matrix method. The left-hand side of the system can be introduced as a multiplication of a tridiagonal matrix and a vector of unknown coefficients. This method is implemented in the {\tt{mpspline}} package  \citep{Ruiz_Jose2022} using the {\tt{scipy}} function, which makes the computation faster. We use this mean-preserving spline to obtain the time series of averaged or middle-term actions and frequencies from eight piecewise functions (\ref{eq:mean_actfreq}). Further, we will use these averaged variables without any indices: $\bm{J} = (J_R, J_z^{max},J_z^{min}, L_z)$, $\bm{\Omega} = (\textcolor{black}{\Omega_R}, \textcolor{black}{\Omega_z}^{max},  \textcolor{black}{\Omega_z}^{min}, \textcolor{black}{\Omega_\varphi})$.   The angles are obtained from the integral $\bm{\theta}(t) = \int_{t_0}^{t} \bm{\Omega}(t)dt + \bm{\theta}_0$, where $t_0$ is the time of the first apocentre passage  for the radial angle and $\theta_R(t_0)=0, \theta_\varphi(t_0) = \varphi(t_0)$ (the azimuthal angle of the first apocentre). The vertical angles at time of first $z$-maximum and minimum are equals $\theta_z^{max}(t_0) = 0$, $\theta_z^{min}(t_0) = \pm\pi$,  respectively.  The sign of $\theta_z^{\text{min}}$ (i.e., $+\pi$ or $-\pi$) depends on whether a $z$-maximum or a $z$-minimum is encountered first.
Finally, the vertical middle-term actions, frequencies, and angles are determined as follows: $J_z = (J_z^{max} + J_z^{min})/2$, $\textcolor{black}{\Omega_z}= (\textcolor{black}{\Omega_z}^{max} + \textcolor{black}{\Omega_z}^{min})/2$, $\theta_z = (\theta_z^{max} + \theta_z^{min})/2$. Thus, we calculate the middle-term action-angle variables close to resonant ones. The left panel of Fig. \ref{fig:aver_actions} demonstrates the orbit in the analysed $N$-body model. The middle and the right ones show the evolution of instantaneous, piecewise, averaged and secular (see next subsection) actions $J_R$ and $J_z$. As expected, the averaged actions smooth out the short-term fluctuations present in the instantaneous values.
\par
Other values we need to estimate are the time derivatives of actions $\dot{\bm{J}}$. Since we use the quadratic spline to find smooth functions of actions $\bm{J}$, their time derivatives are the linear spline. To obtain the smooth functions for $\dot{\bm{J}}$, we use the procedure described above (Eq.~\ref{eq:spline0}-\ref{eq:spline1}), but with different time intervals and input values. Let the time intervals $[t_j, t_{j+1}]$ and values $\dot{y}_j$ be defined as follows:

\begin{equation}
     t_j = \dfrac{t_{i+1} - t_i}{2},\;\;\dot{y}_j = 2\dfrac{y_{i+1} - y_{i}}{t_{i+2} - t_i}\;:\;j = 1, 2,...,n-2,
\end{equation}
here, $y_i$ and $t_i$ have the same meaning as in Eq.~(\ref{eq:spline0})-(\ref{eq:spline1}).
Thus, we numerically evaluate the derivative of the function and relate it to the time interval between the midpoints of the initial ones. Subsequently, we apply a mean-preserving spline once more to achieve smooth temporal dependencies.
\par
However, several issues can arise when calculating averaged action-angle variables. One problem occurs near the boundaries of the spline function. The values may vary significantly, especially when attempting to evaluate a quantity outside the interpolation range, i.e., before $t_1$ or after 
$t_{n+1}$
  (for example, before the first or after the last apocentre). Two approaches can mitigate this problem: (1) truncating all values outside the interval by assigning a \texttt{nan} value before $t_{n-1}$ or $t_n$; (2) assuming that, since the self-consistent $N$-body model starts in equilibrium, the derivative at $t=0$ (or $t_0$) equals zero $S'(t_{0}) =0$, and  $y_{-1} = y_{0}$.  Our implementation provides three options, including two truncation schemes and an additional left boundary condition. The most reliable approach involves integrating the orbit in both the initial and final potentials (with a fixed pattern speed $\Omega_p$) in reverse time order, and then applying the first truncation method.
\par
Another issue is that the spline interpolation is not constrained by the data range. This can occasionally result in negative values for actions such as $J_R$ and $J_z$. Although the \texttt{mpspline} package includes options to address this, it increases computational time by approximately  $10-20$ times. Therefore, we accept that the procedure may, in rare instances, yield unexpected values.
\par
Finally, the last issue relates to orbit types that exhibit rapid changes in the radial period $T_{R,i} = t_{R,i+1} - t_{R,i}$.  This notably affects central orbits and orbits near the ultraharmonic resonance $(\textcolor{black}{\Omega_\varphi} - \textcolor{black}{\Omega_R}/4=\Omega_p)$, which have low eccentricity. For example, some orbits can quickly transit from one apocentre to the next, then spend significantly more time before reaching the subsequent apocentre. This behaviour lead to spline interpolation artefacts, producing large outliers in the frequencies $\textcolor{black}{\Omega_R}$ and $\Omega$. To address this, we merge two consecutive apocentres into a single passage time, defined as $(t_{R,i} +t_{R,i+1})/2$. This merging is applied when the time interval between apocenters satisfies $T_{R,i}<1.4 \cdot T_{R,i+1}$ and  $T_{R,i}<1.4 \cdot T_{R,i-1}$, and when the eccentricity calculated from the pericentre and apocentre distances as $e=2(r_a - r_p)/(r_a+r_p)$ is less than $0.1$. The factors $1.4$ and the eccentricity cutoff were empirically determined through extensive numerical experiments. Although this procedure is approximate, it enables a reasonable estimation of frequencies without significant errors.
\par
Finally, the resulting averaged action-angle variables and frequencies effectively describe motion on medium-term scales. Far from resonances, these quantities remain approximately constant and correspond to trajectories defined by the averaged Hamiltonian (\ref{eq:H_aver}) near a single resonance (Appendix~\ref{ap:aver}  and Section~\ref{sec:PP_bar}), or by Hamiltonians of the form (\ref{eq:H_bar}) in the presence of two fast angles.

\subsubsection{Secular variables}

To characterise long-term orbital evolution on timescales exceeding the libration or circulation periods, we introduce secular actions and frequencies. These quantities are computed in the same way as the medium-term vertical action $J_z$.
The averaged actions and frequencies are further averaged separately between their maxima and minima and then smoothed using a mean-preserving spline. Consequently, two spline functions are obtained for each action and frequency, and the secular variable is defined as the mean of these two functions. Our approach also allows for more precise estimation of integrals of motion, such as the adiabatic invariant $J_v = J_R + J_z+L_z/2$ or the precession rate $\textcolor{black}{\Omega_\varphi} - \textcolor{black}{\Omega_R}/2$ for the bar particles. For this purpose, we apply the same averaging procedure to the corresponding sums of medium-term variables, but only between the maxima and minima of the medium-term angular momentum $L_z$. 
\par
Near the resonance described by Eq.~(\ref{eq:H_pend}), the secular actions and frequencies for librating particles should coincide with the exact resonant action or frequency ($J_{3,\mathrm{res}}$). However, in the case of the more general resonant Hamiltonian Eq.~(\ref{eq:H_aver}) or in scenarios involving two resonances as in (\ref{eq:H_bar}), this equivalence may not hold. Nonetheless, these secular variables continue to characterise the long-term evolution of the particle.
\par
It should be noted that naive averaging over a time window may not account for individual orbital variations and can lead to aliasing artefacts when there is a discrepancy between the libration/circulation period and the width of the time window. If necessary, secular variables can be averaged out over a long enough period of time. \textcolor{black}{This is because} they keep the information about the mean value and are free from major large-scale oscillations.
\par
Additionally, by analysing the time series of medium-term actions and frequencies, one can numerically determine the times of passage through their maxima and minima. This enables estimation of the amplitude and period of variations in the actions and frequencies. Moreover, the Lynden-Bell derivative \citep{Lynden-Bell1979, Zozulia_etal2024a} can be estimated as $\Delta(\textcolor{black}{\Omega_\varphi} - \textcolor{black}{\Omega_R}/2)/\Delta L_z$, where difference $\Delta$ obtain in the time interval between the maximum and minimum of $L_z$. \citet{Zozulia_etal2024a} analysed this value to highlight the bar particles.
\par
Since secular variables are free from the dominant medium-term oscillations, they can be averaged again over an appropriate time window. However, one must remember that long-term trends are orbit-dependent and can change rapidly and that these variables are not entirely free from all medium-term fluctuations. Despite this caveat, such averaging can be useful for a more detailed analysis of orbital behaviour and for identifying subtle structures in action–frequency space. We refer to these averaged quantities as ``Mean Secular'' variables and provide an example of their computation for the $N$-body model in Sec~\ref{sec:act_freq_dift}.

\subsection{Orbit classification}
\label{sec:orb_class}

The galaxy, and the bar in particular, exhibits a rich resonant structure. Each resonance is characterised by the resonant frequency $\Omega_{\mathrm{res}} = \bm{n}\cdot \bm{\Omega}$ and the corresponding resonant angle $\theta_{\mathrm{res}} = \bm{n}\cdot \bm{\theta}$. \textcolor{black}{As we mentioned in Section~\ref{sec:act-ang-nonsym}, this resonant angle can either librate with a near-zero angular velocity or circulate with a positive or negative angular velocity.}
\par
Aside from chaotic regions, each of these modes localises distinct areas within the galaxy phase space. Particles in chaotic trajectories or within evolved potentials can transit between these modes \citep{Chiba_etal2021}. Consequently, particles may be in the process of passing from circulation with positive angular velocity to circulation with negative angular velocity.
\par
\cite{Zozulia_etal2024b} and \cite{Zozulia_etal2025} described how to separate orbits of individual particles based on their resonant averaged angle behaviour into four types. Below, we present the final version of our classification.

\begin{enumerate}
    
 \item Positive Circulation: The resonant angle $\theta_\mathrm{res}$ circulates with a positive angular velocity ($\dot{\theta}_\mathrm{res}>0$), progressing sequentially through values that are integer multiples of $\pi$ in ascending order.

 \item Negative Circulation: The resonant angle $\theta_\mathrm{res}$ circulates with a negative angular velocity ($\dot{\theta}_\mathrm{res}<0$), evolving through values that are integer multiples of $\pi$ in a descending sequence.

 \item Libration around a fixed point:  The resonant angle $\theta_\mathrm{res}$ librates around a stable fixed equilibrium point, which has an angle value $\theta_\mathrm{eq}=0$ or $\pi$, within a range of $\pm \pi$. Determining this regime involves three steps. On the first one, the librating regime is identified when the resonant angle completes at least one full oscillation around the stable point. In practice, this means $\theta_\mathrm{res}$ must cross the equilibrium value at least three times. Let us denote the times of these intersections as $(t_1,\,t_2,...,t_n)$.
 On the second and third steps, the times when the particle entered and left resonance are determined. The particle is considered to enter the resonance at a time $t_{\mathrm{in}}<t_1$. This time is defined by the condition  $|\theta_{\mathrm{res}}(t_{\mathrm{in}}) - \theta_\mathrm{eq}| = \Delta \theta_\mathrm{max}$, where $\Delta \theta_\mathrm{max} = \max{\{|\theta_{\mathrm{res}}(t) - \theta_\mathrm{eq}| \}}$, for $t \in [t_1, t_2]$. Practically, 
$t_{\mathrm{in}}$ is identified as the moment  immediately before $t_1$ when the deviation equalled this maximum amplitude. The resonance exit time 
$t_\mathrm{out}$ is determined by an analogous condition applied after the final crossing $t_n$. The global maximum angular amplitude on the libration can be determined as ${|\theta_{\mathrm{res}}(t) - \theta_\mathrm{eq}| \}}$, for $t \in [t_1, t_n]$. 

 \item Resonance passage (Swing-by): This transient regime is characterised by a temporary capture where the resonant angle $\theta_\mathrm{res}$ crosses the equilibrium angle $\theta_\mathrm{eq}$ only twice. Here, the trajectory approaches the resonance region in phase space but does not complete a full libration cycle, resulting in a reversal of the circulation direction. The entry and exit times of this resonant interaction are defined similarly to the libration case.
\end{enumerate}

The simple set of conditions described above enables the rapid determination of orbital modes influenced by the resonant angle. Furthermore,   in contrast to frequency analysis methods \citep{Portail_etal2015a, Parul_etal2020, Sellwood_Gerhard2020, Smirnov_etal2021, Beraldo2023}, this angular-based approach allows for the localised characterisation of chaotic orbital behaviour. It directly provides estimates for the time spent in resonance and the angular libration amplitude. This advantage stems from its freedom from the fixed time-window constraints inherent in conventional frequency-based techniques.

\subsection{Phase-portrait fitting}
\label{sec:pp_fit}

To investigate the phase space structure of a galaxy, it is essential to understand its geometric configuration. Studying the global shape of phase space provides more comprehensive insights than analysing individual orbits, as it allows one to infer the orbital structures responsible for specific galactic features. For instance, near the corotation resonance with a fixed Jacobi integral, the phase space closely resembles that of a pendulum, and its transformation in this context allows for an explanation of \textcolor{black}{halo--bar dynamical friction and radial migration of orbits \citep{Tremaine_Weinberg_1984, Chiba_etal2021}}.
\par
One of the most commonly employed techniques for exploring the structure of phase space is the Poincaré section. However, this method is subject to a significant limitation: the system must be two-dimensional or reducible to a system with two actions and two angles. Such conditions are satisfied in \textcolor{black}{three-dimensional} axisymmetric galaxies characterised by  ($J_R, J_z, \theta_R, \theta_z$)  or in any two-dimensional \textcolor{black}{planar} disc galaxy exhibiting \textcolor{black}{non-axisymmetric} features such as bars or spiral arms.
\par
To investigate the phase-space structure of a three-dimensional system, we need to reduce the system to the two-dimensional (Sec.~ \ref{sec:H_bar}) one or even one-dimensional system. In the last case, it becomes even easier to study it because it does not contain secular resonances and chaotic regions. The
general form of the Hamiltonian can be expressed by Eq.~\ref{eq:H_2d}. The functions $h(J)$ can be represented as a polynomial from the square of the action $\sum_{k=1}^{k_{max}}{a_k J^{k/2}}$. We consider this polynomial representation because in the epicycle approximation $R - R_0 \propto \sqrt{J_R}$ ($R_0$ is a radius of the guiding centre) and $z \propto \sqrt{J_z}$. There are two notable cases of systems that can be described by this kind of Hamiltonian:  i) orbits in and near the two-dimensional bar $H_{\mathrm{2d}} (J_R, 2\theta_\phi - \theta_R)$ and ii) \textcolor{black}{bar-aligned 3D orbits} $H_{\mathrm{2d}} (J_z, \theta_z - \theta_R)$. In both cases, the orbits correspond to a constant Jacobi integral. Moreover, using the variables $(J_z, \theta_z - \theta_R)$ it is possible to reduce a two-dimensional axisymmetric model to a one-dimensional one. Below, we introduce a method for determining the coefficients of the polynomials described above.
\par
First, a set of orbits is constructed. Their initial conditions in the two-dimensional bar (with its major axis aligned along the $x$-axis) are chosen such that $x=0$, $v_y=0$, and a fixed value of the Jacobi integral $H_J$ is specified. The coordinate $y$ is varied, and the velocity component $v_x$ is determined as a root of the equation $H_J = H(x=0, y, v_x, v_y=0)$, where $H$ denotes the Hamiltonian of the two-dimensional system (\ref{eq:H_gal}).
\par
To investigate the vertical structure of the bar, orbits aligned with the major axis of the bar are selected. These orbits are found in a frozen rotating potential described by (\ref{eq:H_gal}) with initial conditions $(x, y=0, z=z_0, v_x=0, v_y=H_J(x,z_0), v_z=0)$, corresponding to $(\theta_\mathrm{R}=0, \theta_z=0, \theta_\phi=0)$. Here, the Jacobi integral $H_J$ is fixed, and the orbit set is constructed with different values of $z = z_0$. The aim is to determine the value of $x$ such that the apocentres of the orbits lie \textcolor{black}{ in the plane that contains the $z$-axis and the major axis} of the bar. To achieve this, the orbit is integrated over a given time interval, and the sum of the squares of the $y$-coordinate at the apocentres, $\sum y^2(t_\mathrm{apo})$, is minimised.
\par
Second, the specific averaged action-angle $(J_i, \theta_i)$ variables and their time derivative $(\dot{J}_i, \dot{\theta}_i)$ are calculated (Sec. \ref{sec:aver_act-ang}). The Hamiltonian (\ref{eq:H_2d}) represents time derivative from the Hamilton's equations:

\begin{equation}
    \dot{J}(J,\theta) = -\dfrac{\partial H }{\partial\theta};\;\;\; \dot{\theta}(J,\theta) = -\dfrac{\partial H}{\partial J}.
\end{equation}

Both functions are expressed as polynomials with coefficients to be determined. To obtain these coefficients, we minimise the function:

\begin{equation}
    S = \sum_{i}{ \left( \dfrac{\dot{J}(J_i, \theta_i) - \dot{J}_i}{2\sqrt{J_i}\sqrt{J_{max}}} \right)^2 + \left( w_{\theta} \sqrt{J_i} \dfrac{ \dot{\theta}(J_i, \theta_i) - \dot{\theta}_i}{2\pi} \right)^2}.
    \label{eq:cost}
\end{equation}

 This particular form of the loss function is chosen to balance the contributions of derivatives with respect to both the action and the angle. In the expression~(\ref{eq:cost}), $J_{\max}$ denotes the maximal value $J$ across all orbits in the ensemble and $w_\theta$ is a weighting factor introduced to fine-tune the contribution from the angular derivative. Typically, $w_\theta$ is set to unity; variations in this parameter generally have a negligible effect on the optimisation results.

\section{Astrophysical applications}
\label{sec:applications}

We implemented methods from the Section \ref{sec:methods} in the python package \texttt{galport} (the GALactic phase-PORTrate investigator). 
Below, we apply the described methods to the $N$-body model with the mature grown bar.

\subsection{Numerical model}

We use one $N$-body model of a galaxy from \citet{Smirnov_Sotnikova2018}. Initially, the system contains two components: an exponential stellar disc and an isotropic spherical dark halo whose density follows the Navarro-Frenk-White profile~\citep{NFW}. The parameters of the model are expressed in natural units with the gravitational constant set to $G=1$, the disc mass normalised to $M_\mathrm{d}=1$ and the initial exponential scale length of the disc set to $R_\mathrm{d}=1$. Under these scalings, one time unit corresponds approximately to $13.8$ Myr, assuming a disc mass 
$M_\mathrm{d}=5 \times 10^{10} M_\odot$
 and a scale length $R_\mathrm{d}=3.5$ kpc. The initial vertical thickness of the disc is $z_\mathrm{d} = 0.05$, which is incorporated into the vertical density profile using a $\mathrm{sech}^2$ law. Within $4R_\mathrm{d}$, the dark matter halo mass is approximately $M_\mathrm{h}(R < 4R_\mathrm{d}) \approx 1.5 M_\mathrm{d}$ and the minimal Toomre stability parameter reaches $Q(2R_\mathrm{d}) = 1.2$.
\par
A self-consistent evolution of the model is traced up to the time $t=600$ with minimal time resolution $0.125$. The bar forms by the moment $t=80$ and almost immediately starts to grow vertically.  Between moments $150$ and $200$ a buckling event occurs. This leads to the formation of a boxy/peanut-shaped bulge (B/PS). After this process, the bar continues to slowly grow in the radial direction, captures new particles from the disc, and grows vertically as a result of vertical resonance trapping and passage.
\par
\begin{figure}
    \centering
    \includegraphics[width=1.0\linewidth]{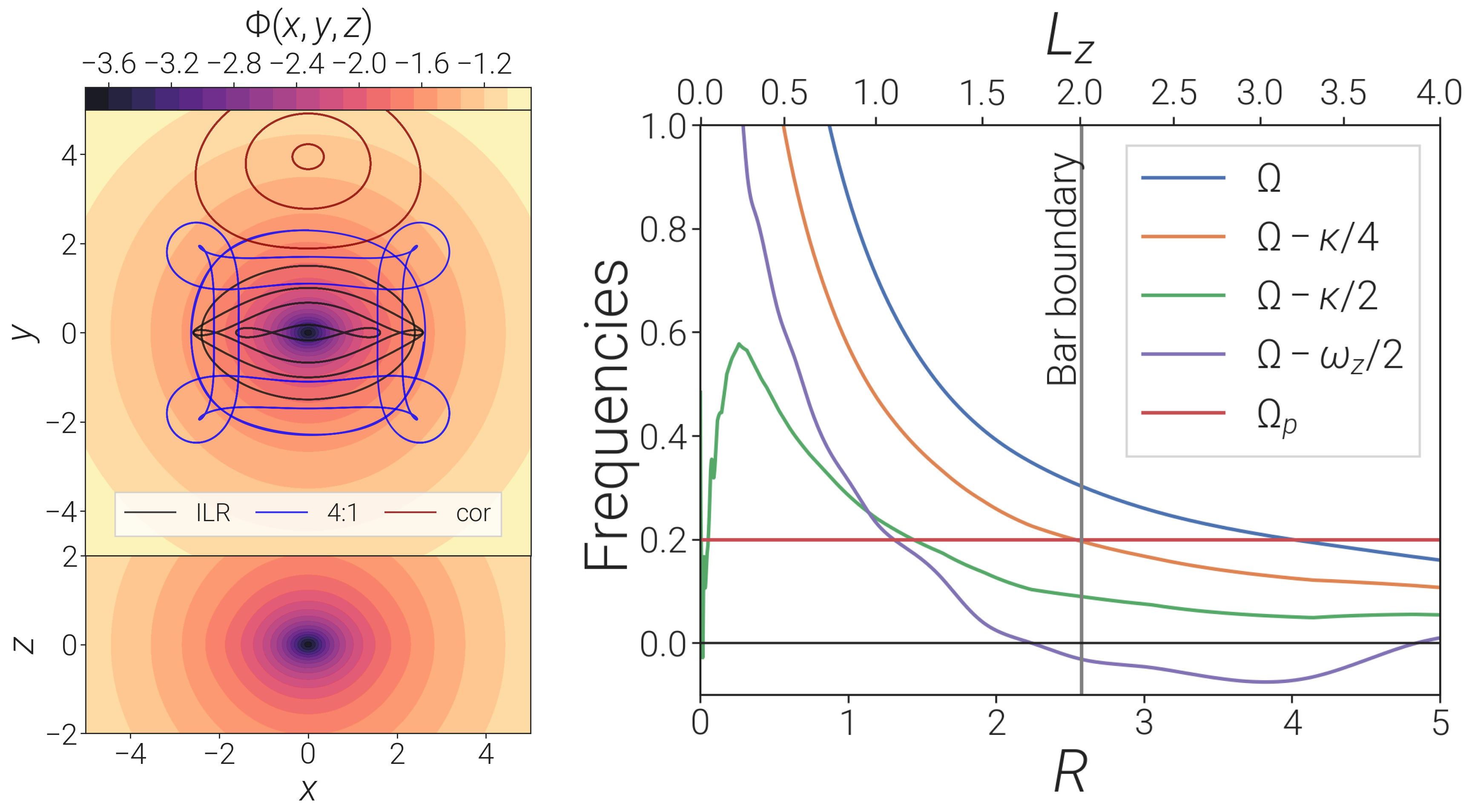}
    \caption{\textit{Left panel}: The potential of the $N$-body model  at time moment $t=400$ on the $xy$ and $xz$ plane. The major axis of a bar aligns along  the $x$ axis. Examples of pure resonant orbits in the inner \textcolor{black}{Lindblad} resonance (ILR, x1), in the ultraharmonic resonance (4:1) and in corotation resonance (CR) are presented on the $xy$-plane.  All these orbits are found in the triaxial potential.  \textit{Right panel:} The dependence of frequencies combinations on the radius in the epicyclic approximation for the axisymmetric part of $N$-body model potential. The red line corresponds to the bar angular speed. The bar boundary is determined by the maximum apocentre of x1 orbits and is shown by a grey vertical line.}
    \label{fig:model}
\end{figure}

 
In this work, we focus on the mature bar stage after the buckling. Specifically, we demonstrate the main results obtained by our package at the time point $t = 400$. To investigate the model, we construct its potential using the \texttt{AGAMA} package. To calculate instantaneous actions at every time with a step of 0.125, we use the multipole expansion of the potential up to order $l=12$ in the meridional angle and leave only the axisymmetric part of the potential (an order $m=0$ in the azimuthal angle). If we investigate the potential of the model in detail at a specific time moment, we represent its full potential as the sum of two components: a disc potential obtained by expanding in cylindrical harmonics   (\texttt{CylSpline}) and a dark halo potential expanded in spherical harmonics (\texttt{Multipole}).  The potential of our model is demonstrated on the left side of Fig. \ref{fig:model}. In the case where the potential is considered in the $xy$-plane, it is assumed to be triaxial symmetric (\texttt{symmetry='Triaxial'} in \texttt{AGAMA}). We find the pure resonant orbits (x1, the ultraharmonic or  4:1 and corotation) in such potential and calculate the mean actions for a set of orbits in these three resonances. Their examples we show in the left panel of Fig.~\ref{fig:model}. The actions are plotted on the plane $(J_R, L_z)$ in Fig. \ref{fig:J_dif_t}. \textcolor{black}{To find pure resonant orbits, we fix the initial $y$ coordinate, set $x=0$ and $v_y=0$, and vary $v_x$ to minimise the residuals of a given resonance condition. For instance, for corotation pure resonance orbits, the sum of $x^2$ at the apocentres and pericentres is minimised (as they should lie along the minor axis of the bar).}
\par
The axisymmetric part of the potential can be used to calculate frequencies using the epicycle approximation.  This approach, however, is not well-suited for the non-axisymmetric potential, as most orbits are far from circular. Nevertheless, we calculate these frequencies in order to compare them with the actual ones and to determine the limitations of this approximation. The right panel of Fig. \ref{fig:model} shows the dependence of the combination of epicyclic frequencies on radius $R$ and on the circular angular momentum  $L_z=\Omega(R)R^2$.  The red line in Fig.~\ref{fig:model} corresponds to the pattern speed $\Omega_p$, the grey line determines the bar boundary from pure resonant x1-orbits. We can see that the bar stretches to an ultraharmonic resonance $\textcolor{black}{\Omega_\varphi}-\textcolor{black}{\Omega_R}/4=\Omega_p$. This is a consequence of the fact that family 4:1 inherits from x1 through the flat orbits with $J_R \approx 0$, where the epicycle approximation should hold (see the left panel of Fig. \ref{fig:model} and Fig. \ref{fig:J_dif_t}).

\subsection{Actions and frequencies on different time scales}
\label{sec:act_freq_dift}

The primary objective of the \texttt{galport} package is to trace orbital evolution across different time scales in action-angle space. To demonstrate its functionality, we compute the actions, angles, and frequencies for every particle in the $N$-body model described above. We find the time series of averaged and secular actions $(J_R, J_z, L_z)$ and frequencies $(\textcolor{black}{\Omega_R}, \textcolor{black}{\Omega_z}, \textcolor{black}{\Omega_\varphi})$ and set of averaged angles $(\theta_R, \theta_z, \theta_\phi)$.  Moreover, the algorithm estimates the period of libration and circulation by the maximum of averaged actions. However, in this paper, we will focus on considering actions and frequencies at a specific time moment $t=400$. Other variables require additional interpretation, which is beyond the scope of this methodological article.
\par
\textcolor{black}{The algorithm takes the coordinates for the time interval from 0 to 600 with a time step of 0.125 and calculates the time series of the variables. A single core of an AMD Ryzen 9 7950X processes 100 orbits in 2.1 seconds. One can also use the built-in parallelisation to speed up the computations or implement a custom workflow by running different orbits on different cores.}
\par
\begin{figure*}
    \centering
    \includegraphics[width=1.0\linewidth]{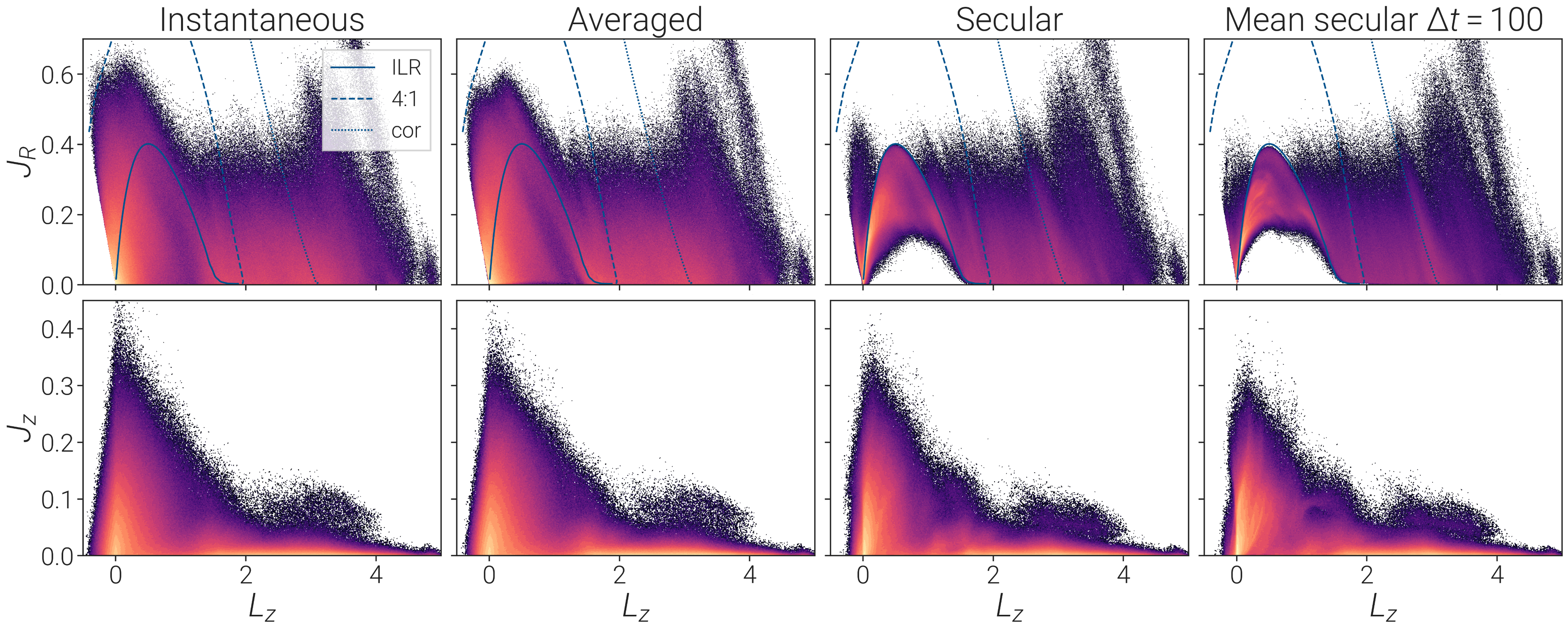}
    \caption{Two-dimensional distributions of the $N$-body galaxy in the action space at the moment $t=400$. They obtained on different time scales: instantaneous or agama actions, medium-term or averaged, long-term or secular, and secular actions averaged between $350$ and $450$ time moments (mean secular). The distributions are shown on the plane $(L_z, J_R)$ for the \textit{top panels} and on the plane $(L_z, J_z)$ for \textit{the bottom} ones.
    The blue solid, dashed and dotted lines on the top panels demonstrate the location of pure resonance orbits (ILR, ultraharmonic 4:1 and corotation) in the action space.}
    \label{fig:J_dif_t}
\end{figure*}

The distributions of action on planes $(L_z, J_R)$ and $(L_z, J_z)$ calculated on different time scales are shown in Fig. \ref{fig:J_dif_t}. We can see that the overall distributions of averaged and instantaneous actions are practically indistinguishable. This is because small-scale fluctuations are smoothed out on such distributions. The distributions of the secular (long-term) actions demonstrate many structures related to resonances. The bar in the $(L_z, J_R)$ plane lies under the envelope corresponding to the pure x1 orbits. In addition, in the disc area, we can trace structures associated with corotation and ultraharmonic resonances. These secular actions can be averaged because they conserve their mean value and are free from large-scale oscillations. This averaging process occurs between $t=350$ and $t=450$, with a time resolution of 5 time units. This means secular actions are plotted on the right panels of Fig.\ref{fig:J_dif_t}. They show the emergence of more subtle structures in the bar, corresponding to different families of orbits. Although their interpretation required additional investigation, the appearance of such structures already shows how well the averaging algorithm works.
\par
\begin{figure*}
    \centering
    \includegraphics[width=1.0\linewidth]{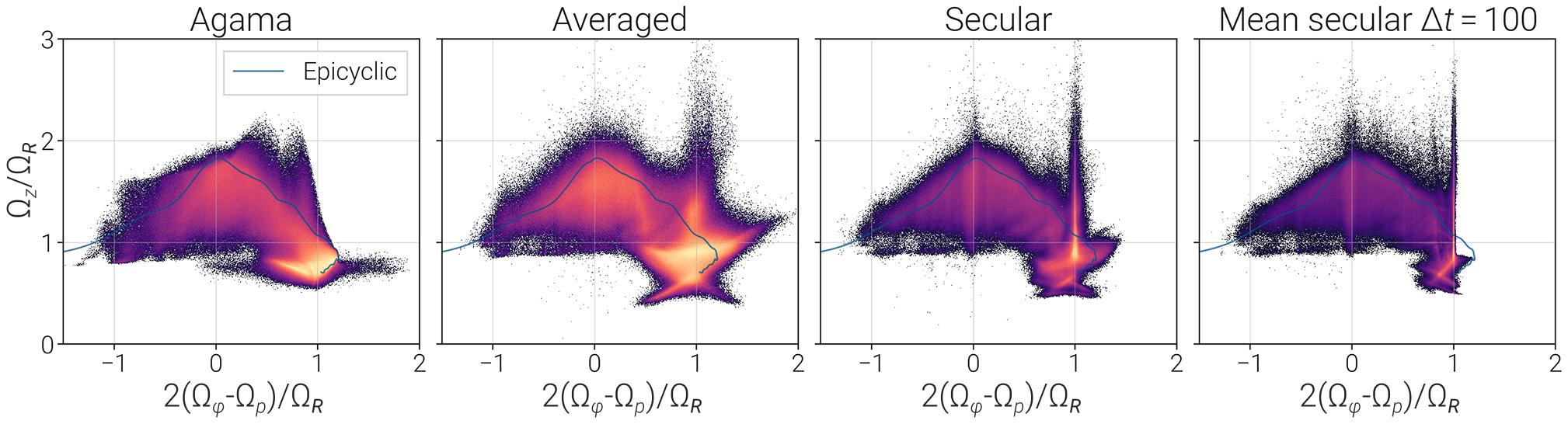}
    \caption{Two-dimensional distributions of $N$-body galaxy in space of frequencies fraction $(2(\textcolor{black}{\Omega_\varphi}-\Omega_p)/\textcolor{black}{\Omega_R}, \textcolor{black}{\Omega_z}/\textcolor{black}{\Omega_R})$. Frequencies are calculated for the same time scales as in Fig. \ref{fig:J_dif_t}. The blue lines correspond to epicycle approximations of frequencies.}
    \label{fig:freq_dif_t}
\end{figure*}

The distributions of frequencies on the plane  $(2(\textcolor{black}{\Omega_\varphi}-\Omega_p)/\textcolor{black}{\Omega_R}, \textcolor{black}{\Omega_z}/\textcolor{black}{\Omega_R})$ are shown in Fig.\ref{fig:freq_dif_t}. The left plot demonstrates frequencies calculated in axisymmetric potential by \texttt{AGAMA} package. The other three panels correspond to the same time scales as in Fig.~\ref{fig:J_dif_t}. If we compare the axisymmetric and averaged frequencies, the former generally reproduces the frequency distribution in the disc ($2(\textcolor{black}{\Omega_\varphi - \Omega_p})/\textcolor{black}{\Omega_R} < 1$), but poorly describes the bar region ($2(\textcolor{black}{\Omega_\varphi - \Omega_p})/\textcolor{black}{\Omega_R} \approx 1$). We discuss the reason for this in Sec.~\ref{sec:inst_act}.The secular frequencies enable more accurate tracing of features in both the disc and the bar. There are other methods for estimating frequencies; for example, frequency analysis \citep{Parul_etal2020, Smirnov_etal2021, Beraldo2023} and the least squares method for azimuthal angle \citep{Ceverino_Klypin2007}. However, the main advantage of our method is its ability to directly track the temporal evolution of frequencies across different timescales.  This is because our method outputs a continuous time series rather than a single value. Moreover, if a more accurate analysis is required, we can again average the secular frequencies (see the right panel of Fig. \ref{fig:freq_dif_t}). This approach makes it possible to resolve subtler structures in the frequency ratio space in our model.

\begin{figure*}
    \centering
    \includegraphics[width=1.0\linewidth]{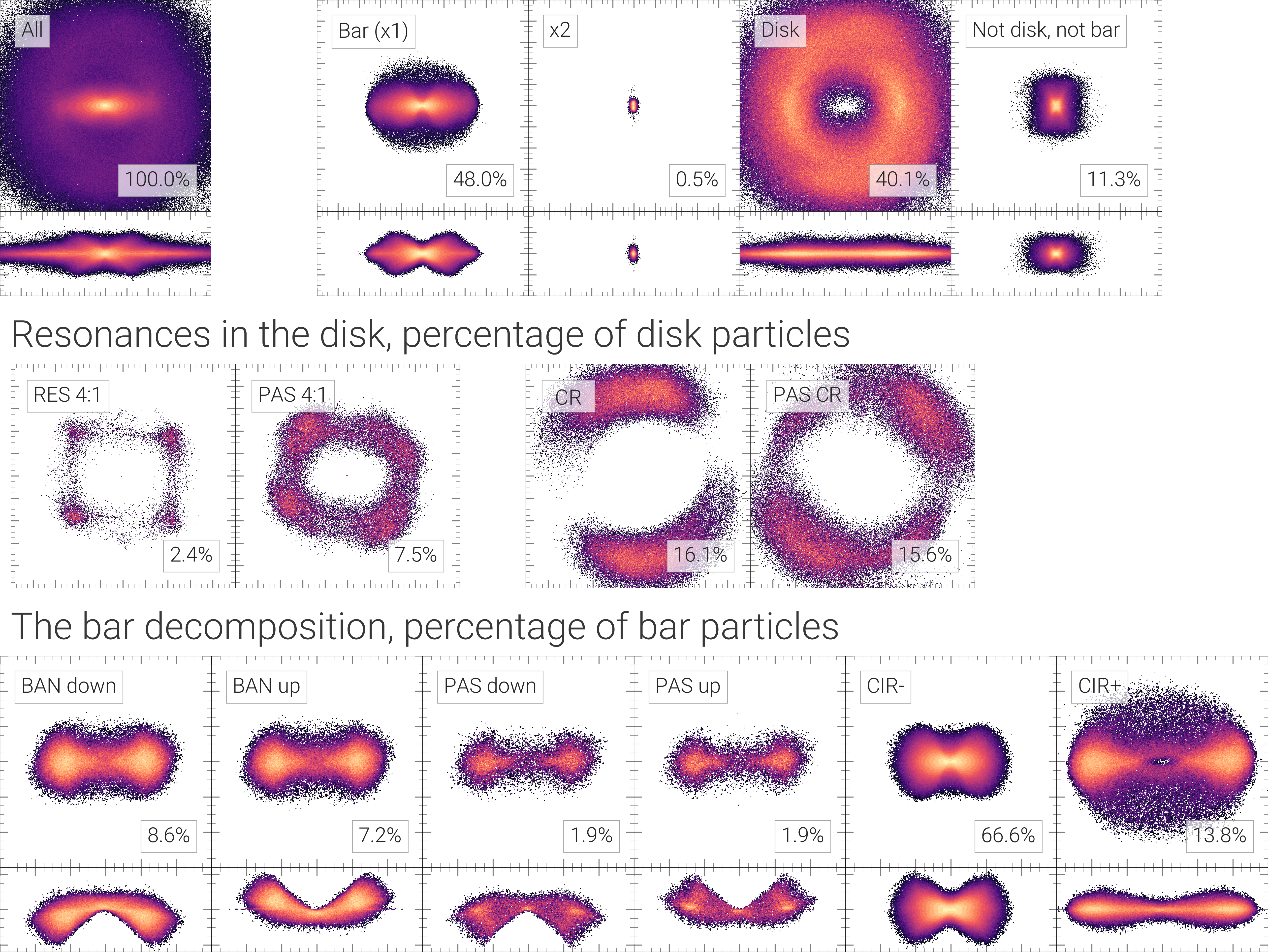}
    \caption{The dynamic decomposition of the $N$-body model. \textit{Upper row, left direction}: All model orbits, the bar orbits ($2\theta_\phi - \theta_R$ librates around $0$), orbits near x2 ($2\theta_\phi - \theta_R$ librate around $0$), particles that do not belong to the bar or disc and disc particles. \textit{Central row}: particles in ultraharmonic resonance ($4\theta_\phi - \theta_R$ librates around the $0$ or $\pi$ or pass through these angles) and corotation  particles ($2\theta_\phi$ librates around the $\pi$ or pass through these angles).  \textit{Lowest row}: The bar particles with different behaviours of resonant angle $\theta_z - \theta_R$.}
    \label{fig:gal_decompose}
\end{figure*}

\subsection{Dynamic decomposition}
\label{sec:ang_decompose}

The dynamic decomposition is one of the  sophisticated problems while exploring the galaxy model. Usually decomposition is based on the frequencies \citep{Portail_etal2015a, Parul_etal2020}. In Sec.~\ref{sec:orb_class} we introduce how the method of dynamic decomposition based on the resonant angle behaviour works. 
\par
In Fig.~\ref{fig:gal_decompose} we illustrate the decomposition of $N$-body model (the snapshot is provided on the top left corner). In the top panel, we present a base decomposition of the galaxy by the four components:
\begin{enumerate}
    \item Bar (x1): the resonant angle $2\theta_\phi - \theta_R$ librates around $0$. Orbits are aligned with the major axis of the bar, including orbits with negative $\Omega$. Approximately half of all particles are confined in the bar. 
    \item x2: the resonant angle $2\theta_\phi - \theta_R$ librates around $\pi$. Orbits align to the minor axis of the bar. Only $0.5\%$ of the model orbits exhibits this behaviour.
    \item Not disc, not bar: Particular to this model, it is the central component of not librating orbits near the bar that have left the disc plane. This component is defined by three circulation criteria: 1) $2\dot{\theta}_\phi - \dot{\theta}_R<0$; 2) $4\dot{\theta}_\phi - \dot{\theta}_R>0$; 3) $\dot{\theta}_z - \dot{\theta}_R >0$. This component can also be selected from not bar orbits by the boundary value of the secular $L_z$ \citep{Zozulia_etal2025}. It should be noted that in other models, these orbits may have other angular behaviour, e.g., circulating orbits with $2\dot{\theta}_\phi - \dot{\theta}_R>0$ can be added. \textcolor{black}{The nature of this component requires further research. As can be seen, these particles form a central, extended, vertical, and non-spherical structure. Some of these particles are likely chaotic, switching between circulation and libration, while others are resonant but not associated with the ILR. They can also be identified in the $(2(\textcolor{black}{\Omega_\varphi - \Omega_p})/\textcolor{black}{\Omega_R}, \textcolor{black}{\Omega_z}/\textcolor{black}{\Omega_R})$ plane. }
    \item Disc: All other particles. 
\end{enumerate}
\par
In the middle row of Fig.~\ref{fig:gal_decompose} we also select the orbits in ultraharmonic (4:1) and corotation resonances and passage through them. As expected, most of the orbits in the 4:1 resonance form a quadrangular structure and constitute a relatively small fraction of the disc particles (about $10\%$). Corotation orbits account for approximately one-third of all disc particles and are located perpendicular to the bar. In addition, resonance orbits are slightly shifted counterclockwise relative to the minor axis of the bar, while passage orbits are shifted clockwise. This is consistent with the findings described in \cite{Chiba_etal2021} and is connected with the bar slowdown.
\par
 The lower row of Fig.~\ref{fig:gal_decompose} demonstrates the vertical dynamics decomposition of the bar, as performed in \citet{Zozulia_etal2025}. We selected orbits, the angle $\theta_z - \theta_R$ of which circulate with positive and negative speed, librate around $0$ (BAN up) or $\pi$ (BAN down) or pass through this angle (PAS up and down). This decomposition allows us to separate the different vertical components of the bar (thin, thick or intermediate). In ~\cite{Zozulia_etal2025}, we analysed the fraction of these components during the buckling process.
\par
To create this decomposition, we use the time series of angles $(\theta_R, \theta_z, \theta_\phi)$ from $t=50$ to $t=550$ with a time step of 5 time units.  Note that the procedure does not require high-resolution time series. This reflects the fact that the libration and/or circulation period is several times longer than the radial oscillation period of a particle on an orbit. However, the numerical values of the angles themselves must be obtained by processing a time series, the accuracy of which allows for the precise determination of the time of passage of the apocentre. \textcolor{black}{The orbital classification is a computationally efficient procedure; given already calculated angles, it takes approximately 4.5 minutes for $4\times10^6$ orbits.}

\subsection{Phase portraits of the bar}
\label{sec:PP_bar}

\begin{figure*}
    \centering
    \includegraphics[width=1.0\linewidth]{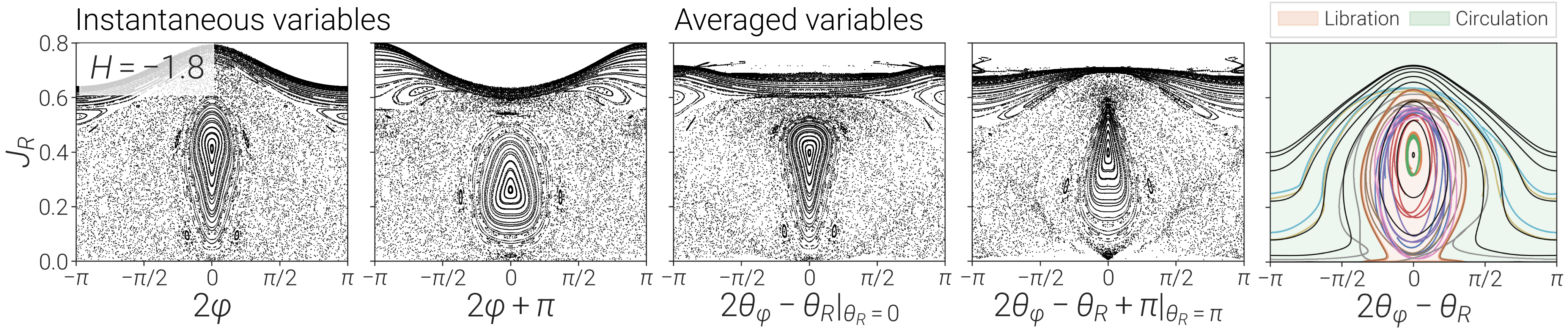}
    \caption{The structure of the action-angle space $(J_R, 2\theta_\phi - \theta_R)$ for orbits in the disc plane with the fixed Jacobi integral of $-1.8$. \textit{Four left panels} demonstrate Poincaré maps. Points correspond to apocentres ($\theta_R =0$) and pericentres ($\theta_R = \pi$) for instantaneous and averaged variables. \textit{Right panel:}  Coloured lines correspond to the evolution of average actions and angles of some orbits. Black lines correspond to isolines of the Hamiltonian $H_{\mathrm{2d}}(J_R, 2\theta_\phi - \theta_R)$ (\ref{eq:H_2d}), found using the phase-portrait fitting procedure from Sec. \ref{sec:pp_fit}. \textcolor{black}{The orange line separates the libration area (orange background) from the circulation region (green background)}.}
    \label{fig:PP_bar}
\end{figure*}

\begin{figure}
    \centering
    \includegraphics[width=1\linewidth]{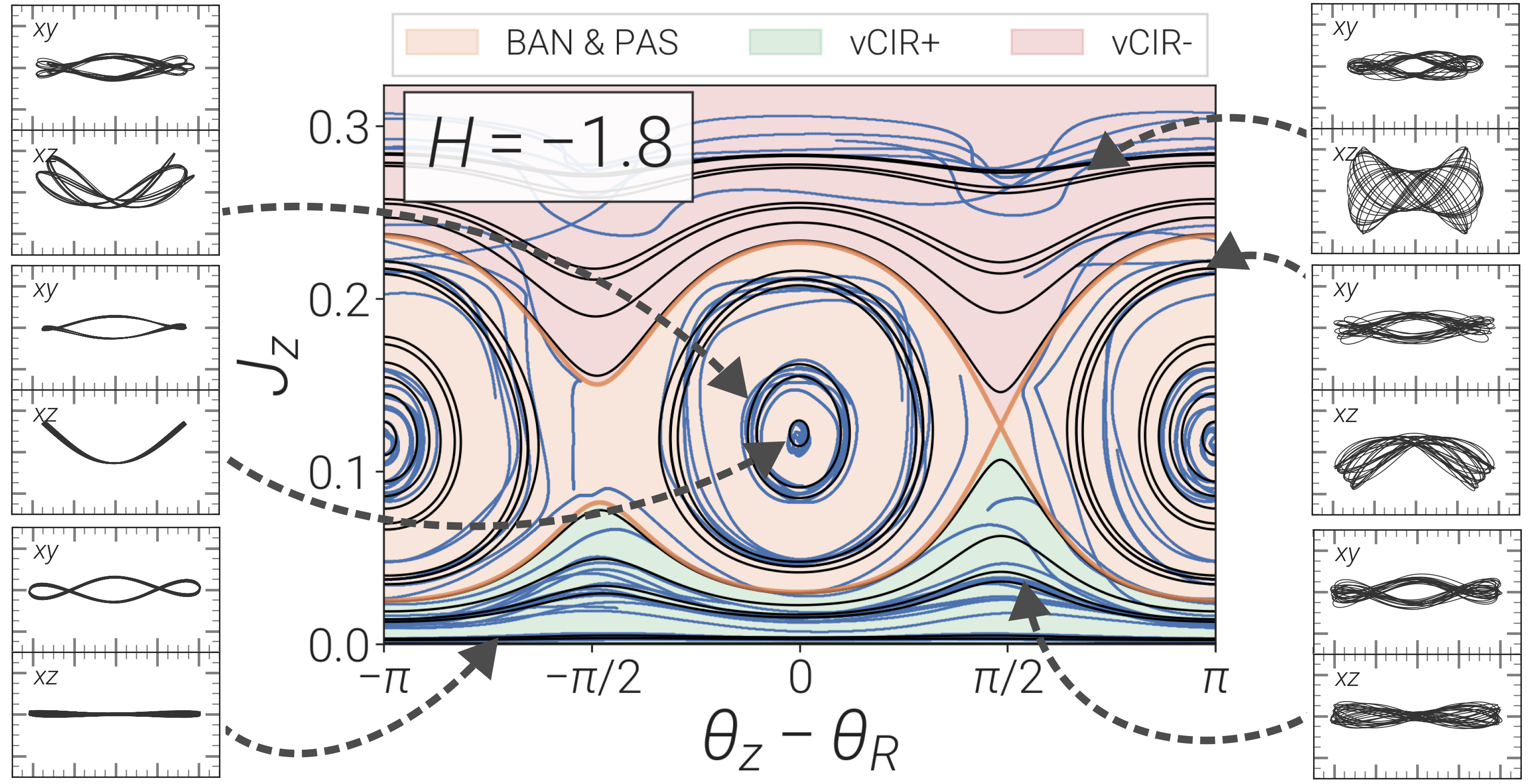}
    \caption{The evolution of \textcolor{black}{bar-aligned 3D orbits} on the plane $(\theta_z-\theta_R, J_z$ (blue lines). The Jacobi integral equals $-1.8$ for all of them. Examples of some orbits are shown on $xy$ and $xz$ planes. Both face-on and edge on views  are displayed in the square $(-2.5,2.5)\times(-1.5,1.5)$. The black lines on the central plot correspond to isolines of the two-dimensional fitted Hamiltonian $H_{\mathrm{2d}}(\theta_z-\theta_R, J_z)$  (Sec. \ref{sec:pp_fit}). \textcolor{black}{The orange line separates the libration area (orange background) from the circulation regions, where $\dot{\theta}_z - \dot{\theta}_R > 0$ (green background) and $\dot{\theta}_z - \dot{\theta}_R < 0$ (red background), respectively.}
   }
    \label{fig:PP_ma}
\end{figure}

In Sec.~\ref{sec:pp_fit} we describe the algorithm for how to obtain the approximated analytical two-dimensional Hamiltonian $H_{\mathrm{2d}}(J, \theta)$ of the bar orbits in the disc plane and align along its major axis. This procedure should help to simplify the study of the structure of the phase space. In particular, the analytic Hamiltonian helps to estimate the position of the resonances and \textcolor{black}{separatrices} and the period of libration and circulation for orbits with different initial $(J, \theta)$. 
\par
First, we consider orbits in the disc plane within a frozen potential rotating with a constant pattern speed. For the orbital integration, we use the \texttt{AGAMA} package ~\citep{agama}. The classical method for investigating the phase space of these orbits is the Poincaré map. In Fig.~\ref{fig:PP_bar} we demonstrate these maps for both instantaneous and averaged action-angle variables on the time scales of $2000$ time units for orbits with a fixed Jacobi integral of $-1.8$. The sections are obtained in apocentres and pericentres. We use a simple azimuthal angle as an instantaneous angular variable. Noticeably, these maps reflect perfectly the phase space of these orbits.
\textcolor{black}{They may even be better than} the classical sections on the $(x, v_x)$ plane for $y=0$. \textcolor{black}{The reason is that they show the weight of the libration area by azimuthal angle and action $J_R$ of the regular tori (possibly plot the same maps for $z$-component of the angular momentum $L_z$)}. Note that although averaged actions are obtained as a spline smoothing of instantaneous variables, they also can be used for plotting the Poincaré maps. \textcolor{black}{In particular, they separate the regular and chaotic orbits. This means that they continue to be phase variables. However, in contrast to instantaneous variables, the averaged ones} do not alter the position of the fixed point and generally show smaller changes from pericentre to apocentre.
The regular tori definitely undergo reshaping, but we believe that this is related to the influence of secular resonances between the libration period and the radial oscillation period.
\par
The right panel of Fig.~\ref{fig:PP_bar} shows the behaviour of the averaged action-angle variables $(J_R, 2\theta_\phi - \theta_R)$ over \textcolor{black}{$100$} time units and the results of the Hamiltonian fitting procedure. Several features are evident in this plot: (1) the represented orbits complete several revolutions or circulations; (2) regular and chaotic orbits are faintly discernible on these timescales; and (3) the isolines of the fitted Hamiltonian generally follow the averaged $J_R$ and $2\theta_\phi - \theta_R$, capturing the main features of the phase space, including the stationary point and separatrix. Note that the galaxy's potential and pattern speed can change significantly over $200$ time units (about $2.8$ billion years). To this end, one could use such two-dimensional Hamiltonians calculated at different time moments to track the \textcolor{black}{the phase space evolution in the $(J_R, 2\theta_\phi - \theta_R)$ plane --- specifically, to estimate the boundaries of libration and circulation regions (as shown in Fig.~\ref{fig:PP_bar}).} However, such a study lies beyond the scope of this paper.
\par
The next example of the Hamiltonian fitting procedure is demonstrated in the Fig.~\ref{fig:PP_ma}. This represents the evolution of \textcolor{black}{bar-aligned 3D orbits} on a plane $(J_z, \theta_z-\theta_R)$ with the fixed value of the Jacobi integral ($H=-1.8$). In the plot, we can see the overall structure of the phase space, including the areas where the angle  $\theta_z - \theta_R$ increases (beyond the lower separatrix) and decreases (above the upper separatrix). There are also two libration regions associated with banana orbits (as shown by the examples of orbits). The two-dimensional fitted Hamiltonian $H_{\mathrm{2d}}(J_z, \theta_z-\theta_R)$ captures the main features of this phase space and can be used to approximately trace the general orbital evolution in this space.
\par
\textcolor{black}{The computational time required for the phase-portrait fitting depends on both the number of orbits used and the number of its parameters. For instance, the fitting procedure alone for the 10 orbits shown in Fig.~\ref{fig:PP_bar} takes 15 seconds, while for Fig.~\ref{fig:PP_ma} it takes 5 seconds (on an AMD Ryzen 9 7950X). The preliminary search for near-bar orbits is almost instantaneous; in contrast, identifying a set of 10 bar-aligned 3D orbits requires calculating 4000 orbits. This search takes 15 seconds on 16 cores of an AMD Ryzen 9 7950X and approximately 100 seconds on 12 cores of an Intel Core i7-1260P.}

\section{Conclusion}
\label{sec:conclusion}
In this work we present the \texttt{galport} \texttt{Python} package. It is designed for analysing disc galaxy simulations in action-angle space. The primary problem the package solves is estimating time series of action-angle variables from time series of Cartesian coordinates and velocities on different time scales. We identify two key dynamical time scales: a medium-term scale (associated with radial or vertical oscillations) and a long-term scale (associated with the libration or circulation period near a resonance). The variables derived on these scales are termed average and secular, respectively. Analysing the average angle enables orbital classification relative to resonances. The package also provides tools for investigating bar potential and deriving analytical expressions for the one-dimensional Hamiltonian that describes phase space in the disc plane (in variables $J_R$ and $2\theta_\phi - \theta_R$) and along the bar's major axis  ($J_z$ and $\theta_z - \theta_R$).
\par
We have demonstrated the package's capabilities by applying it to a typical $N$-body model of a galaxy with a mature, slowly evolving bar. The key results of this analysis are:
\begin{enumerate}
    \item Action and Frequency Distributions: We obtained the distribution of all model particles in action space  $(L_z, J_R)$ and $(L_z, J_z)$, and in the space of frequency ratios $(2(\textcolor{black}{\Omega_\varphi - \Omega_p})/\textcolor{black}{\Omega_R}, \textcolor{black}{\Omega_z}/\textcolor{black}{\Omega_R})$ for instantaneous, averaged and secular variables. Crucially, the frequency distributions demonstrate that instantaneous frequencies calculated in an axisymmetric approximation are unsuitable for analysing bar orbits.
    \item Robust Orbital Classification: The package enabled a detailed dynamical decomposition of the model. We identified and quantified the primary components: bar particles ($48\%$), the disc ones ($40\%$), central particles not associated with either the bar or the disc ($11\%$) and a small population of orbits perpendicular to the bar (x2 orbits, $0.5\%$). We also separately examined resonant orbits within the disc and the vertical behaviour of bar particles  (see  Sec.~\ref{sec:orb_class} and Fig.~\ref{fig:gal_decompose} for details).
    \item Phase-Space Structure of the Bar: As presented in Sec.~\ref{sec:PP_bar}, the analytical description of the bar's phase space clearly captures its structure for a fixed Jacobi integral. This approach condenses the behaviour of many orbits into a single schematic, revealing the underlying phase-space geometry.
\end{enumerate}

As a result, the package \texttt{galport} allow us to simplify the interpretation of the result of galactic simulation through the action-angle space.
\par
Future development will expand the package's functionality for more detailed and fast model analysis, to include: parallel computation for large orbit sets, tools for isolating pure resonant orbits, and phase-space analysis near other resonances (e.g., corotation and outer Lindblad resonances).

\section*{Acknowledgements}

I am grateful to N.J.Sotnikova, A.A. Smirnov, V.S. Kostiuk for their valuable comments on this paper. \textcolor{black}{I thank the anonymous reviewer whose valuable recommendations allowed me to significantly improve the manuscript}. Financial support was provided by the Russian Science Foundation, grant no. 25-22-00738.
This work could not be possible without the follow \texttt{python} packages: \texttt{AGAMA} \citep{agama},   \texttt{mpsplines} \citep{Ruiz_Jose2022}, \texttt{numpy} \citep{numpy} and \texttt{scipy} \citep{scipy}.


\appendix

\section{Averaging}
\label{ap:aver}

The Hamiltonian of a disc galaxy can be expressed as a function of three actions and angles $H(\bm{J}, \bm{\theta}) = H_0(\bm{J}) + H_1(\bm{J}, \bm{\theta})$, where $\bm{J} = (J_R, J_z, L_z)$ and $\bm{\theta} = (\theta_R, \theta_z, \theta_\phi)$ (see Sec.~\ref{sec:act-ang-sym}).
Let us consider the motion of a particle in the action-angle space near two resonances \textcolor{black}{$\bm{N}_1 \cdot \bm{\Omega}$ and $\bm{N}_2 \cdot \bm{\Omega}$, where $\bm{\Omega} = \partial H_0 / \partial \bm{J}$}. In this case, we can introduce the new resonance action-angle variables,  which are related linearly with the initial ones. The canonical transformation between new and initial variables can be written as:
\begin{equation}
\bm{\theta}' 
= A \bm{\theta} =
\begin{pmatrix}
\bm{N}_0  \\
\bm{N}_1\\
\bm{N}_2
\end{pmatrix}  \bm{\theta} =
\begin{pmatrix}
\theta_0 \\
\theta_1 \\
\theta_2
\end{pmatrix};\;\;
\bm{J}'
= (A^{-1})^T \bm{J} = \begin{pmatrix}
J_0 \\
J_1 \\
J_2
\end{pmatrix},
\end{equation}
here we assume that the angle $\theta_0$ takes the values of either $\theta_R$ or $\theta_z$ and evolves fast, while the other two angles evolve slowly. In these new variables, the perturbed part of the Hamiltonian~(\ref{eq:pert_H}) can be separated into two parts, one associated with slow and fast angles:
\begin{equation}
\label{eq:ap_H}
    H(\bm{J}, \bm{\theta}) = H_0(\bm{J}) +\sum_{\bm{n}\in\bm{m}} h_{\bm{n}}(\bm{J}) e^{i\bm{n}\cdot\bm{\theta}} +\sum_{\bm{n}\notin\bm{m}} h_{\bm{n}}(\bm{J}) e^{i\bm{n}\cdot\bm{\theta}},
\end{equation}
where $\bm{m} = m_1\bm{N}_1 + m_2\bm{N}_2$ and $m_1,\,m_2$ are integers. The last term depends only on the fast angle $\theta_0$ and can be excluded after averaging by it. In this case, the averaging Hamiltonian takes the following form: 
\begin{equation}
    \overline{\bm{H}} (\bm{J}', \theta_1, \theta_2) = H_0(\bm{J}') +  \sum_{\bm{m}} h_{\bm{m}}(\bm{J}') e^{i(m_1\theta_1 + m_2\theta_2)},
\end{equation}
Solving Hamilton's equations for this system yields the evolution of the averaged actions  $\bm{J}'=(J_0=const, J_1(t), J_2(t))$ and two slow angles $\theta_1(t), \theta_2(t)$. The evolution of the remaining fast angle is found from $\dot\theta_0(t) = \partial \overline{\bm{H}}(J_0, J_1(t), J_2(t)) /  \partial J_0$. Then, we return to the original action-angle variables $\bm{J}^a = A^T\bm{J'}$. Here the new notation means that we do not take into account only the non-resonant terms. Let us return to the original Hamiltonian (Eq.~\ref{eq:ap_H}) to estimate the error introduced by averaging the initial variables $(\bm{J}, \bm{\theta})$ (as described in Sec.~\ref{sec:aver_act-ang}). For this aim, we use the first-order perturbation expressions  (Eq.~(\ref{eq:pert_theta})-(\ref{eq:nonres_J})). The averaging time $T$ is defined by the conditions $\theta_0(0) = 0$ and $\theta_0(T) = 2\pi$. The error of this procedure for frequencies and actions is given by the second term in the right-hand part of the following equations:

\begin{equation}
\label{eq:ap_integrals}
\begin{split}
    \left< \dot{\bm{\theta}} \right> &= \left< \dot{\bm{\theta}}^a \right> + \dfrac{1}{T} \int_0^T \sum_{\bm{n} \notin \bm{m}} \dfrac{\partial h_{\bm{n}}(\bm{J}^a)}{\partial \bm{J}^a} e^{i\bm{n} \cdot \bm{\theta}^a} dt; \\
    \left< \bm{J}\right> &= \left< \bm{J}^{a} \right> - \dfrac{1}{T} \int_0^T {\sum_{\bm{n} \notin \bm{m}}{\dfrac{\bm{n} \cdot h_{\bm{n}}(\bm{J}^a)}{\bm{n}\cdot\bm{\Omega}(\bm{J}^a)} e^{i\bm{n} \cdot \bm{\theta}^a}} dt}.
\end{split}
\end{equation}

Since the frequencies $\dot{\bm{\theta}}^a$ and actions $\bm{J}^a$ change slowly, only the complex exponential in the sums varies significantly over the averaging interval. Let us again return to the resonance angles:
\begin{multline}
\bm{n} \cdot \bm{\theta}^a = n_1\theta^a_R + n_2 \theta^a_z+n_3 \theta^a_\phi = \bm{n} \cdot A^{-1}\bm{\theta}' = \bm{l}\cdot\bm{\theta}'  \approx \\ \left( l_0\left< \dot{\theta_0}\right> + l_1\left< \dot{\theta_1}\right> + l_2\left<\dot{\theta_2}\right> \right) t + \bm{l} \cdot \Delta \bm{\theta}',
\end{multline}
here $\Delta \bm{\theta}'$ is the initial phase of resonant angles. Then, with $T \approx 2\pi/ \left< \dot{\theta}_0 \right>$, the integrals in Eq.~(\ref{eq:ap_integrals}) can be evaluated:

\begin{equation}
\begin{split}
    \left< \dot{\bm{\theta}} \right> &= \left< \dot{\bm{\theta}}^a \right> -
\\
    &\dfrac{i}{2\pi}  \sum_{\bm{l}:l_1\neq 0} \dfrac{\partial h_{\bm{l}A}(\bm{J}^a)}{\partial \bm{J}^a} \dfrac{ e^{i \bm{l}\cdot \bm{\theta}' (T)} - e^{i \bm{l}\cdot \bm{\theta}' (0)}}{l_0 + l_1 \left< \dot{\theta}_1 \right> / \left< \dot{\theta}_0 \right> + l_2 \left< \dot{\theta}_2 \right> / \left< \dot{\theta}_0 \right>};
\\
    \left< \bm{J}\right> &= \left< \bm{J}^{a} \right> +
\\
    &\dfrac{i}{2\pi} \sum_{\bm{l}:l_1\neq 0}{\dfrac{\bm{l}A \cdot h_{\bm{l}A}(\bm{J}^a)}{\bm{l}\cdot A\bm{\Omega}(\bm{J}^a)} \dfrac{ e^{i \bm{n}\cdot \bm{\theta}' (T)} - e^{i \bm{n}\cdot \bm{\theta}' (0)}}{l_0 + l_1 \left< \dot{\theta}_1 \right> / \left< \dot{\theta}_0 \right> + l_2 \left< \dot{\theta}_2 \right> / \left< \dot{\theta}_0 \right>}}.
\end{split}
\end{equation}

The denominator in these expressions is large, because $\dot{\theta_0} \gg \dot{\theta}_1, \dot{\theta}_2$. A key feature is that terms with $l_1 = l_2 = 0$ vanish, as the numerators in both expressions become zero.  Thus, the influence of the non-resonant terms is greatly suppressed, demonstrating that the averaged variables estimate the dynamics of the averaged Hamiltonian system. We note that even in systems with no slow angles or only one slow angle, this averaging procedure reduces the influence of fast terms by a factor of at least $2\pi$. In such cases, the averaged variables $(\bm{J}^a, \bm{\theta}^a)$ are either constants or are governed by single-resonance dynamics (see Sec.~\ref{sec:act-ang-nonsym}).

















\bibliographystyle{cas-model2-names}

\bibliography{cas-refs}

@ARTICLE{Ansar_2025,
       author = {{Ansar}, Sioree and {Pearson}, Sarah and {Sanderson}, Robyn E. and {Arora}, Arpit and {Hopkins}, Philip F. and {Wetzel}, Andrew and {Cunningham}, Emily C. and {Quinn}, Jamie},
        title = "{Bar Formation and Destruction in the FIRE-2 Simulations}",
      journal = {\apj},
         year = 2025,
        month = jan,
       volume = {978},
       number = {1},
          eid = {37},
        pages = {37},
          doi = {10.3847/1538-4357/ad8b45},
archivePrefix = {arXiv},
       eprint = {2309.16811},
 primaryClass = {astro-ph.GA},
       adsurl = {https://ui.adsabs.harvard.edu/abs/2025ApJ...978...37A}
}

@ARTICLE{Arora_etal_2022,
       author = {{Arora}, Arpit and {Sanderson}, Robyn E. and {Panithanpaisal}, Nondh and {Cunningham}, Emily C. and {Wetzel}, Andrew and {Garavito-Camargo}, Nicol{\'a}s},
        title = "{On the Stability of Tidal Streams in Action Space}",
      journal = {\apj},
         year = 2022,
        month = nov,
       volume = {939},
       number = {1},
          eid = {2},
        pages = {2},
          doi = {10.3847/1538-4357/ac93fb},
archivePrefix = {arXiv},
       eprint = {2207.13481},
 primaryClass = {astro-ph.GA},
       adsurl = {https://ui.adsabs.harvard.edu/abs/2022ApJ...939....2A}
}

@BOOK{Arnold1978,
       author = {{Arnold}, Vladimir Igorevich},
        title = "{Mathematical methods of classical mechanics}",
         year = 1978,
       adsurl = {https://ui.adsabs.harvard.edu/abs/1978mmcm.book.....A}
}

@ARTICLE{Beraldo2023,
       author = {{Beraldo e Silva}, Leandro and {Debattista}, Victor P. and {Anderson}, Stuart Robert and {Valluri}, Monica and {Erwin}, Peter and {Daniel}, Kathryne J. and {Deg}, Nathan},
        title = "{Orbital Support and Evolution of Flat Profiles of Bars (Shoulders)}",
      journal = {\apj},
         year = 2023,
        month = sep,
       volume = {955},
       number = {1},
          eid = {38},
        pages = {38},
          doi = {10.3847/1538-4357/ace976},
archivePrefix = {arXiv},
       eprint = {2303.04828},
 primaryClass = {astro-ph.GA},
       adsurl = {https://ui.adsabs.harvard.edu/abs/2023ApJ...955...38B}
}

@ARTICLE{Binney_McMillan2011,
       author = {{Binney}, James and {McMillan}, Paul},
        title = "{Models of our Galaxy - II}",
      journal = {Monthly Notices of the Royal Astronomical Society},
         year = 2011,
        month = may,
       volume = {413},
       number = {3},
        pages = {1889-1898},
          doi = {10.1111/j.1365-2966.2011.18268.x},
archivePrefix = {arXiv},
       eprint = {1101.0747},
 primaryClass = {astro-ph.GA},
       adsurl = {https://ui.adsabs.harvard.edu/abs/2011MNRAS.413.1889B}
}

@ARTICLE{Binney2012,
author = {{Binney}, James},
title = "{Actions for axisymmetric potentials}",
journal = {Monthly Notices of the Royal Astronomical Society},
year = 2012,
month = oct,
volume = {426},
number = {2},
pages = {1324-1327},
doi = {10.1111/j.1365-2966.2012.21757.x},
archivePrefix = {arXiv},
eprint = {1207.4910},
primaryClass = {astro-ph.GA},
adsurl = {https://ui.adsabs.harvard.edu/abs/2012MNRAS.426.1324B}
}

@ARTICLE{Binney_McMillan_2016,
       author = {{Binney}, James and {McMillan}, Paul J.},
        title = "{Torus mapper: a code for dynamical models of galaxies}",
      journal = {Monthly Notices of the Royal Astronomical Society},
         year = 2016,
        month = feb,
       volume = {456},
       number = {2},
        pages = {1982-1998},
          doi = {10.1093/mnras/stv2734},
archivePrefix = {arXiv},
       eprint = {1511.07754},
 primaryClass = {astro-ph.GA},
       adsurl = {https://ui.adsabs.harvard.edu/abs/2016MNRAS.456.1982B}
}

@ARTICLE{Binney_2018,
author = {{Binney}, James},
title = "{Orbital tori for non-axisymmetric galaxies}",
journal = {Monthly Notices of the Royal Astronomical Society},
year = 2018,
month = feb,
volume = {474},
number = {2},
pages = {2706-2724},
doi = {10.1093/mnras/stx2835},
archivePrefix = {arXiv},
eprint = {1710.11360},
primaryClass = {astro-ph.GA},
adsurl = {https://ui.adsabs.harvard.edu/abs/2018MNRAS.474.2706B}
}

@ARTICLE{Binney_2020b,
       author = {{Binney}, James},
        title = "{Trapped orbits and solar-neighbourhood kinematics}",
      journal = {Monthly Notices of the Royal Astronomical Society},
         year = 2020,
        month = jun,
       volume = {495},
       number = {1},
        pages = {895-904},
          doi = {10.1093/mnras/staa1103},
archivePrefix = {arXiv},
       eprint = {1912.07023},
 primaryClass = {astro-ph.GA},
       adsurl = {https://ui.adsabs.harvard.edu/abs/2020MNRAS.495..895B}
}

@ARTICLE{Binney_Vasiliev_2023,
       author = {{Binney}, James and {Vasiliev}, Eugene},
        title = "{Self-consistent models of our Galaxy}",
      journal = {Monthly Notices of the Royal Astronomical Society},
         year = 2023,
        month = apr,
       volume = {520},
       number = {2},
        pages = {1832-1847},
          doi = {10.1093/mnras/stad094},
archivePrefix = {arXiv},
       eprint = {2206.03523},
 primaryClass = {astro-ph.GA},
       adsurl = {https://ui.adsabs.harvard.edu/abs/2023MNRAS.520.1832B}
}

@ARTICLE{Binney_Vasiliev_2024,
       author = {{Binney}, James and {Vasiliev}, Eugene},
        title = "{Chemodynamical models of our Galaxy}",
      journal = {Monthly Notices of the Royal Astronomical Society},
         year = 2024,
        month = jan,
       volume = {527},
       number = {2},
        pages = {1915-1934},
          doi = {10.1093/mnras/stad3312},
archivePrefix = {arXiv},
       eprint = {2306.11602},
 primaryClass = {astro-ph.GA},
       adsurl = {https://ui.adsabs.harvard.edu/abs/2024MNRAS.527.1915B}
}

@ARTICLE{Binney_Spergel1982,
author = {{Binney}, J. and {Spergel}, D.},
title = "{Spectral stellar dynamics}",
journal = {\apj},
year = 1982,
month = jan,
volume = {252},
pages = {308-321},
doi = {10.1086/159559}
}

@BOOK{Binney_Tremaine2008,
author = {{Binney}, J. and {Tremaine}, S.},
title = {{Galactic Dynamics: Second Edition}},
booktitle = {Galactic Dynamics: Second Edition, by James Binney and Scott Tremaine.~ISBN 978-0-691-13026-2 (HB).~Published by Princeton University Press, Princeton, NJ USA, 2008.},
year = {2008},
publisher = {Princeton University Press},
}

@ARTICLE{galpy,
author = {{Bovy}, Jo},
title = "{galpy: A python Library for Galactic Dynamics}",
journal = {\apjs},
year = 2015,
month = feb,
volume = {216},
number = {2},
eid = {29},
pages = {29},
doi = {10.1088/0067-0049/216/2/29},
archivePrefix = {arXiv},
eprint = {1412.3451},
primaryClass = {astro-ph.GA},
adsurl = {https://ui.adsabs.harvard.edu/abs/2015ApJS..216...29B}
}

@ARTICLE{Ceverino_Klypin2007,
author = {{Ceverino}, D. and {Klypin}, A.},
title = "{Resonances in barred galaxies}",
journal = {Monthly Notices of the Royal Astronomical Society},
eprint = {astro-ph/0703544},
year = 2007,
month = aug,
volume = 379,
pages = {1155-1168},
doi = {10.1111/j.1365-2966.2007.12001.x}
}

@ARTICLE{Chiba_etal2021,
       author = {{Chiba}, Rimpei and {Friske}, Jennifer K.~S. and {Sch{\"o}nrich}, Ralph},
        title = "{Resonance sweeping by a decelerating Galactic bar}",
      journal = {Monthly Notices of the Royal Astronomical Society},
         year = 2021,
        month = jan,
       volume = {500},
       number = {4},
        pages = {4710-4729},
          doi = {10.1093/mnras/staa3585},
archivePrefix = {arXiv},
       eprint = {1912.04304},
 primaryClass = {astro-ph.GA},
       adsurl = {https://ui.adsabs.harvard.edu/abs/2021MNRAS.500.4710C}
}

@ARTICLE{Chiba_etal2021b,
       author = {{Chiba}, Rimpei and {Sch{\"o}nrich}, Ralph},
        title = "{Tree-ring structure of Galactic bar resonance}",
      journal = {Monthly Notices of the Royal Astronomical Society},
         year = 2021,
        month = aug,
       volume = {505},
       number = {2},
        pages = {2412-2426},
          doi = {10.1093/mnras/stab1094},
archivePrefix = {arXiv},
       eprint = {2102.08388},
 primaryClass = {astro-ph.GA},
       adsurl = {https://ui.adsabs.harvard.edu/abs/2021MNRAS.505.2412C}
}

@ARTICLE{Chiba_Schonrich_2022,
       author = {{Chiba}, Rimpei and {Sch{\"o}nrich}, Ralph},
        title = "{Oscillating dynamical friction on galactic bars by trapped dark matter}",
      journal = {Monthly Notices of the Royal Astronomical Society},
         year = 2022,
        month = jun,
       volume = {513},
       number = {1},
        pages = {768-787},
          doi = {10.1093/mnras/stac697},
archivePrefix = {arXiv},
       eprint = {2109.10910},
 primaryClass = {astro-ph.GA},
       adsurl = {https://ui.adsabs.harvard.edu/abs/2022MNRAS.513..768C}
}

@ARTICLE{Chiba_2023,
       author = {{Chiba}, Rimpei},
        title = "{Dynamical friction and feedback on galactic bars in the general fast-slow regime}",
      journal = {Monthly Notices of the Royal Astronomical Society},
         year = 2023,
        month = nov,
       volume = {525},
       number = {3},
        pages = {3576-3596},
          doi = {10.1093/mnras/stad2324},
archivePrefix = {arXiv},
       eprint = {2305.00022},
 primaryClass = {astro-ph.GA},
       adsurl = {https://ui.adsabs.harvard.edu/abs/2023MNRAS.525.3576C}
}

@ARTICLE{Debattista_etal2020,
author = {{Debattista}, Victor P. and {Liddicott}, David J. and {Khachaturyants}, Tigran and {Beraldo e Silva}, Leandro},
title = "{Box/peanut-shaped bulges in action space}",
journal = {Monthly Notices of the Royal Astronomical Society},
year = 2020,
month = nov,
volume = {498},
number = {3},
pages = {3334-3350},
doi = {10.1093/mnras/staa2568},
archivePrefix = {arXiv},
eprint = {1911.01084},
primaryClass = {astro-ph.GA}
}

@ARTICLE{Debattista_etal2025,
       author = {{Debattista}, Victor P. and {Khachaturyants}, Tigran and {Amarante}, Jo{\~a}o A.~S. and {Carr}, Christopher and {Beraldo e Silva}, Leandro and {Laporte}, Chervin F.~P.},
        title = "{Azimuthal metallicity variations, spiral structure, and the failure of radial actions based on assuming axisymmetry}",
      journal = {Monthly Notices of the Royal Astronomical Society},
         year = 2025,
        month = feb,
       volume = {537},
       number = {2},
        pages = {1620-1645},
          doi = {10.1093/mnras/staf035},
archivePrefix = {arXiv},
       eprint = {2402.08356},
 primaryClass = {astro-ph.GA},
       adsurl = {https://ui.adsabs.harvard.edu/abs/2025MNRAS.537.1620D}
}

@ARTICLE{Dillamore_etal2024,
       author = {{Dillamore}, Adam M. and {Belokurov}, Vasily and {Evans}, N. Wyn},
        title = "{Radial halo substructure in harmony with the Galactic bar}",
      journal = {Monthly Notices of the Royal Astronomical Society},
         year = 2024,
        month = aug,
       volume = {532},
       number = {4},
        pages = {4389-4407},
          doi = {10.1093/mnras/stae1789},
archivePrefix = {arXiv},
       eprint = {2402.14907},
 primaryClass = {astro-ph.GA},
       adsurl = {https://ui.adsabs.harvard.edu/abs/2024MNRAS.532.4389D}
}

@ARTICLE{Fardal_2015,
       author = {{Fardal}, Mark A. and {Huang}, Shuiyao and {Weinberg}, Martin D.},
        title = "{Generation of mock tidal streams}",
      journal = {Monthly Notices of the Royal Astronomical Society},
         year = 2015,
        month = sep,
       volume = {452},
       number = {1},
        pages = {301-319},
          doi = {10.1093/mnras/stv1198},
archivePrefix = {arXiv},
       eprint = {1410.1861},
 primaryClass = {astro-ph.GA},
       adsurl = {https://ui.adsabs.harvard.edu/abs/2015MNRAS.452..301F}
}

@ARTICLE{Fragkoudi_etal2025,
       author = {{Fragkoudi}, Francesca and {Grand}, Robert J.~J. and {Pakmor}, R{\"u}diger and {G{\'o}mez}, Facundo and {Marinacci}, Federico and {Springel}, Volker},
        title = "{Bar formation and evolution in the cosmological context: inputs from the Auriga simulations}",
      journal = {Monthly Notices of the Royal Astronomical Society},
         year = 2025,
        month = apr,
       volume = {538},
       number = {3},
        pages = {1587-1608},
          doi = {10.1093/mnras/staf389},
archivePrefix = {arXiv},
       eprint = {2406.09453},
 primaryClass = {astro-ph.GA},
       adsurl = {https://ui.adsabs.harvard.edu/abs/2025MNRAS.538.1587F}
}

@ARTICLE{Ghosh_etal2023,
       author = {{Ghosh}, Soumavo and {Trick}, Wilma H. and {Green}, Gregory M.},
        title = "{Quantifying the influence of bars on action-based dynamical modelling of disc galaxies}",
      journal = {Monthly Notices of the Royal Astronomical Society},
         year = 2023,
        month = jul,
       volume = {523},
       number = {1},
        pages = {991-1008},
          doi = {10.1093/mnras/stad1525},
archivePrefix = {arXiv},
       eprint = {2212.06184},
 primaryClass = {astro-ph.GA},
       adsurl = {https://ui.adsabs.harvard.edu/abs/2023MNRAS.523..991G}
}

@ARTICLE{Hirashima_2025,
       author = {{Hirashima}, Keiya and {Fujii}, Michiko S. and {Saitoh}, Takayuki R. and {Harada}, Naoto and {Nomura}, Kentaro and {Yoshikawa}, Kohji and {Hirai}, Yutaka and {Asano}, Tetsuro and {Moriwaki}, Kana and {Iwasawa}, Masaki and {Okamoto}, Takashi and {Makino}, Junichiro},
        title = "{The First Star-by-star $N$-body/Hydrodynamics Simulation of Our Galaxy Coupling with a Surrogate Model}",
      journal = {arXiv e-prints},
         year = 2025,
        month = oct,
          eid = {arXiv:2510.23330},
        pages = {arXiv:2510.23330},
          doi = {10.48550/arXiv.2510.23330},
archivePrefix = {arXiv},
       eprint = {2510.23330},
 primaryClass = {astro-ph.GA},
       adsurl = {https://ui.adsabs.harvard.edu/abs/2025arXiv251023330H}
}

@ARTICLE{Hunt_etal2019,
       author = {{Hunt}, Jason A.~S. and {Bub}, Mathew W. and {Bovy}, Jo and {Mackereth}, J. Ted and {Trick}, Wilma H. and {Kawata}, Daisuke},
        title = "{Signatures of resonance and phase mixing in the Galactic disc}",
      journal = {Monthly Notices of the Royal Astronomical Society},
         year = 2019,
        month = nov,
       volume = {490},
       number = {1},
        pages = {1026-1043},
          doi = {10.1093/mnras/stz2667},
archivePrefix = {arXiv},
       eprint = {1904.10968},
 primaryClass = {astro-ph.GA},
       adsurl = {https://ui.adsabs.harvard.edu/abs/2019MNRAS.490.1026H}
}

@ARTICLE{Kawata2021,
author = {{Kawata}, Daisuke and {Baba}, Junichi and {Hunt}, Jason A.~S. and {Sch{\"o}nrich}, Ralph and {Ciuc{\u{a}}}, Ioana and {Friske}, Jennifer and {Seabroke}, George and {Cropper}, Mark},
title = "{Galactic bar resonances inferred from kinematically hot stars in Gaia EDR3}",
journal = {Monthly Notices of the Royal Astronomical Society},
year = 2021,
month = nov,
volume = {508},
number = {1},
pages = {728-736},
doi = {10.1093/mnras/stab2582},
archivePrefix = {arXiv},
eprint = {2012.05890},
primaryClass = {astro-ph.GA},
adsurl = {https://ui.adsabs.harvard.edu/abs/2021MNRAS.508..728K}
}

@BOOK{Lichtenberg_Lieberman_1992,
author = {{Lichtenberg}, A. and {Lieberman}, M.},
title = "{Regular and Chaotic Dynamics}",
year = 1992,
adsurl = {https://ui.adsabs.harvard.edu/abs/1992rcd..book.....L}
}

@ARTICLE{Lynden-Bell1979,
author = {{Lynden-Bell}, D.},
title = "{On a mechanism that structures galaxies.}",
journal = {Monthly Notices of the Royal Astronomical Society},
year = 1979,
month = apr,
volume = {187},
pages = {101-107},
doi = {10.1093/mnras/187.1.101}
}

@ARTICLE{Lynden-Bell_Kalnajs1972,
author = {{Lynden-Bell}, D. and {Kalnajs}, A.~J.},
title = "{On the generating mechanism of spiral structure}",
journal = {Monthly Notices of the Royal Astronomical Society},
year = 1972,
month = jan,
volume = {157},
pages = {1},
doi = {10.1093/mnras/157.1.1},
adsurl = {https://ui.adsabs.harvard.edu/abs/1972MNRAS.157....1L}
}

@ARTICLE{Machado_Manos2016,
author = {{Machado}, R.~E.~G. and {Manos}, T.},
title = "{Chaotic motion and the evolution of morphological components in a time-dependent model of a barred galaxy within a dark matter halo}",
journal = {Monthly Notices of the Royal Astronomical Society},
year = "2016",
month = "Jun",
volume = {458},
number = {4},
pages = {3578-3591},
doi = {10.1093/mnras/stw572},
archivePrefix = {arXiv},
eprint = {1603.02294},
primaryClass = {astro-ph.GA}
}

@ARTICLE{Malhan_etal2022,
       author = {{Malhan}, Khyati and {Ibata}, Rodrigo A. and {Sharma}, Sanjib and {Famaey}, Benoit and {Bellazzini}, Michele and {Carlberg}, Raymond G. and {D'Souza}, Richard and {Yuan}, Zhen and {Martin}, Nicolas F. and {Thomas}, Guillaume F.},
        title = "{The Global Dynamical Atlas of the Milky Way Mergers: Constraints from Gaia EDR3-based Orbits of Globular Clusters, Stellar Streams, and Satellite Galaxies}",
      journal = {\apj},
         year = 2022,
        month = feb,
       volume = {926},
       number = {2},
          eid = {107},
        pages = {107},
          doi = {10.3847/1538-4357/ac4d2a},
archivePrefix = {arXiv},
       eprint = {2202.07660},
 primaryClass = {astro-ph.GA},
       adsurl = {https://ui.adsabs.harvard.edu/abs/2022ApJ...926..107M}
}

@ARTICLE{McGill_Binney_1990,
       author = {{McGill}, Colin and {Binney}, James},
        title = "{Torus construction in general gravitational potentials}",
      journal = {Monthly Notices of the Royal Astronomical Society},
         year = 1990,
        month = jun,
       volume = {244},
        pages = {634-645},
       adsurl = {https://ui.adsabs.harvard.edu/abs/1990MNRAS.244..634M}
}

@ARTICLE{Myeong_2018,
       author = {{Myeong}, G.~C. and {Evans}, N.~W. and {Belokurov}, V. and {Sanders}, J.~L. and {Koposov}, S.~E.},
        title = "{The Milky Way Halo in Action Space}",
      journal = {\apjl},
         year = 2018,
        month = apr,
       volume = {856},
       number = {2},
          eid = {L26},
        pages = {L26},
          doi = {10.3847/2041-8213/aab613},
archivePrefix = {arXiv},
       eprint = {1802.03351},
 primaryClass = {astro-ph.GA},
       adsurl = {https://ui.adsabs.harvard.edu/abs/2018ApJ...856L..26M}
}

@ARTICLE{NFW,
author = {{Navarro}, J.~F. and {Frenk}, C.~S. and {White}, S.~D.~M.},
title = "{The Structure of Cold Dark Matter Halos}",
journal = {\apj},
eprint = {astro-ph/9508025},
year = 1996,
month = may,
volume = 462,
pages = {563},
doi = {10.1086/177173}
}

@ARTICLE{Parul_etal2020,
author = {{Parul}, Hanna D. and {Smirnov}, Anton A. and {Sotnikova}, Natalia Ya.},
title = "{Orbital ingredients for cooking X-structures in edge-on galaxies}",
journal = {arXiv e-prints},
year = 2020,
month = feb,
eid = {arXiv:2002.06627},
pages = {arXiv:2002.06627},
archivePrefix = {arXiv},
eprint = {2002.06627},
primaryClass = {astro-ph.GA}
}

@ARTICLE{Polyachenko_Shukhman2020a,
author = {{Polyachenko}, E.~V. and {Shukhman}, I.~G.},
title = "{On the Lynden-Bell Bar Formation Mechanism in Galactic Disks}",
journal = {Astronomy Letters},
year = 2020,
month = jan,
volume = {46},
number = {1},
pages = {12-23},
doi = {10.1134/S1063773719120053},
adsurl = {https://ui.adsabs.harvard.edu/abs/2020AstL...46...12P}
}

@ARTICLE{Portail_etal2015a,
author = {{Portail}, M. and {Wegg}, C. and {Gerhard}, O. and
{Martinez-Valpuesta}, I.},
title = "{Made-to-measure models of the Galactic box/peanut bulge: stellar and total mass in the bulge region}",
journal = {Monthly Notices of the Royal Astronomical Society},
year = "2015",
month = "Mar",
volume = {448},
number = {1},
pages = {713-731},
doi = {10.1093/mnras/stv058},
archivePrefix = {arXiv},
eprint = {1502.00633},
primaryClass = {astro-ph.GA}
}

@ARTICLE{Quillen2002,
author = {{Quillen}, A.~C.},
title = "{Growth of a Peanut-shaped Bulge via Resonant Trapping of Stellar Orbits in the Vertical Inner Lindblad Resonances}",
journal = {\aj},
eprint = {astro-ph/0203170},
year = 2002,
month = aug,
volume = 124,
pages = {722-732},
doi = {10.1086/341753}
}

@ARTICLE{Quillen_etal2014,
author = {{Quillen}, A.~C. and {Minchev}, I. and {Sharma}, S. and {Qin}, Y.-J. and 
{Di Matteo}, P.},
title = "{A vertical resonance heating model for X- or peanut-shaped galactic bulges}",
journal = {Monthly Notices of the Royal Astronomical Society},
archivePrefix = "arXiv",
eprint = {1307.8441},
year = 2014,
month = jan,
volume = 437,
pages = {1284-1307},
doi = {10.1093/mnras/stt1972}
}

@ARTICLE{Reino_etal_2021,
       author = {{Reino}, Stella and {Rossi}, Elena M. and {Sanderson}, Robyn E. and {Sellentin}, Elena and {Helmi}, Amina and {Koppelman}, Helmer H. and {Sharma}, Sanjib},
        title = "{Galactic potential constraints from clustering in action space of combined stellar stream data}",
      journal = {Monthly Notices of the Royal Astronomical Society},
         year = 2021,
        month = apr,
       volume = {502},
       number = {3},
        pages = {4170-4193},
          doi = {10.1093/mnras/stab304},
archivePrefix = {arXiv},
       eprint = {2007.00356},
 primaryClass = {astro-ph.GA},
       adsurl = {https://ui.adsabs.harvard.edu/abs/2021MNRAS.502.4170R}
}

@ARTICLE{Ruiz_Jose2022,
author = {{Ruiz-Arias}, Jos{\'e} A.},
title = "{Mean-preserving interpolation with splines for solar radiation modeling}",
journal = {Solar Energy},
year = 2022,
month = dec,
volume = {248},
pages = {121-127},
doi = {10.1016/j.solener.2022.10.038},
adsurl = {https://ui.adsabs.harvard.edu/abs/2022SoEn..248..121R}
}

@ARTICLE{Sanders_2012,
       author = {{Sanders}, Jason},
        title = "{Angle-action estimation in a general axisymmetric potential}",
      journal = {\mnras},
         year = 2012,
        month = oct,
       volume = {426},
       number = {1},
        pages = {128-139},
          doi = {10.1111/j.1365-2966.2012.21698.x},
archivePrefix = {arXiv},
       eprint = {1208.2813},
 primaryClass = {astro-ph.GA},
       adsurl = {https://ui.adsabs.harvard.edu/abs/2012MNRAS.426..128S}
}

@ARTICLE{Sanders_Binney2015,
author = {{Sanders}, Jason L. and {Binney}, James},
title = "{A fast algorithm for estimating actions in triaxial potentials}",
journal = {Monthly Notices of the Royal Astronomical Society},
year = 2015,
month = mar,
volume = {447},
number = {3},
pages = {2479-2496},
doi = {10.1093/mnras/stu2598},
archivePrefix = {arXiv},
eprint = {1412.2093},
primaryClass = {astro-ph.GA},
adsurl = {https://ui.adsabs.harvard.edu/abs/2015MNRAS.447.2479S}
}

@ARTICLE{Sellwood_Gerhard2020,
author = {{Sellwood}, J.~A. and {Gerhard}, Ortwin},
title = "{Three mechanisms for bar thickening}",
journal = {Monthly Notices of the Royal Astronomical Society},
year = 2020,
month = jul,
volume = {495},
number = {3},
pages = {3175-3191},
doi = {10.1093/mnras/staa1336},
archivePrefix = {arXiv},
eprint = {2005.05184},
primaryClass = {astro-ph.GA},
adsurl = {https://ui.adsabs.harvard.edu/abs/2020MNRAS.495.3175S}
}

@ARTICLE{Smirnov_Sotnikova2018,
author = {{Smirnov}, Anton A. and {Sotnikova}, Natalia Ya},
title = "{What determines the flatness of X-shaped structures in edge-on galaxies?}",
journal = {Monthly Notices of the Royal Astronomical Society},
year = "2018",
month = "Dec",
volume = {481},
number = {3},
pages = {4058-4076},
doi = {10.1093/mnras/sty2423},
archivePrefix = {arXiv},
eprint = {1809.06167},
primaryClass = {astro-ph.GA}
}

@ARTICLE{Smirnov_etal2021,
author = {{Smirnov}, Anton A. and {Tikhonenko}, Iliya S. and {Sotnikova}, Natalia Ya},
title = "{Face-on structure of barlenses and boxy bars: an insight from spectral dynamics}",
journal = {Monthly Notices of the Royal Astronomical Society},
year = 2021,
month = apr,
volume = {502},
number = {4},
pages = {4689-4707},
doi = {10.1093/mnras/stab327},
archivePrefix = {arXiv},
eprint = {2007.09090},
primaryClass = {astro-ph.GA},
adsurl = {https://ui.adsabs.harvard.edu/abs/2021MNRAS.502.4689S}
}

@ARTICLE{Sormani_etal2022,
       author = {{Sormani}, Mattia C. and {Sanders}, Jason L. and {Fritz}, Tobias K. and {Smith}, Leigh C. and {Gerhard}, Ortwin and {Sch{\"o}del}, Rainer and {Magorrian}, John and {Neumayer}, Nadine and {Nogueras-Lara}, Francisco and {Feldmeier-Krause}, Anja and {Mastrobuono-Battisti}, Alessandra and {Schultheis}, Mathias and {Shahzamanian}, Banafsheh and {Vasiliev}, Eugene and {Klessen}, Ralf S. and {Lucas}, Philip and {Minniti}, Dante},
        title = "{Self-consistent modelling of the Milky Way's nuclear stellar disc}",
      journal = {Monthly Notices of the Royal Astronomical Society},
         year = 2022,
        month = may,
       volume = {512},
       number = {2},
        pages = {1857-1884},
          doi = {10.1093/mnras/stac639},
archivePrefix = {arXiv},
       eprint = {2111.12713},
 primaryClass = {astro-ph.GA},
       adsurl = {https://ui.adsabs.harvard.edu/abs/2022MNRAS.512.1857S}
}

@ARTICLE{Smirnov_etal2025,
       author = {{Smirnov}, Anton A. and {Bajkova}, Anisa T. and {Bobylev}, Vadim V. and {Zozulia}, Viktor D.},
        title = "{Exploring the Galactic potential: Insights from cosmological simulations of the Milky Way}",
      journal = {\prd},
         year = 2025,
        month = dec,
       volume = {112},
       number = {12},
          eid = {123523},
        pages = {123523},
          doi = {10.1103/bhk9-63cb},
       adsurl = {https://ui.adsabs.harvard.edu/abs/2025PhRvD.112l3523S}
}

@ARTICLE{Sun_2023,
       author = {{Sun}, GuangChen and {Wang}, Yougang and {Liu}, Chao and {Long}, Richard J. and {Chen}, Xuelei and {Gao}, Qi},
        title = "{Classifying Globular Clusters and Applying them to Estimate the mass of the Milky Way}",
      journal = {Research in Astronomy and Astrophysics},
         year = 2023,
        month = jan,
       volume = {23},
       number = {1},
          eid = {015013},
        pages = {015013},
          doi = {10.1088/1674-4527/ac9e91},
archivePrefix = {arXiv},
       eprint = {2210.12336},
 primaryClass = {astro-ph.GA},
       adsurl = {https://ui.adsabs.harvard.edu/abs/2023RAA....23a5013S}
}

@ARTICLE{Tikhonenko_etal2021,
       author = {{Tikhonenko}, Iliya S. and {Smirnov}, Anton A. and {Sotnikova}, Natalia Ya.},
        title = "{First direct identification of the barlens vertical structure in galaxy models}",
      journal = {\aap},
         year = 2021,
        month = apr,
       volume = {648},
          eid = {L4},
        pages = {L4},
          doi = {10.1051/0004-6361/202140703},
archivePrefix = {arXiv},
       eprint = {2103.02988},
 primaryClass = {astro-ph.GA},
       adsurl = {https://ui.adsabs.harvard.edu/abs/2021A&A...648L...4T}
}

@ARTICLE{Trick_etal2016,
       author = {{Trick}, Wilma H. and {Bovy}, Jo and {Rix}, Hans-Walter},
        title = "{Action-Based Dynamical Modeling for the Milky Way Disk}",
      journal = {\apj},
         year = 2016,
        month = oct,
       volume = {830},
       number = {2},
          eid = {97},
        pages = {97},
          doi = {10.3847/0004-637X/830/2/97},
archivePrefix = {arXiv},
       eprint = {1605.08601},
 primaryClass = {astro-ph.GA},
       adsurl = {https://ui.adsabs.harvard.edu/abs/2016ApJ...830...97T}
}

@ARTICLE{Trick_etal2019,
       author = {{Trick}, Wilma H. and {Coronado}, Johanna and {Rix}, Hans-Walter},
        title = "{The Galactic disc in action space as seen by Gaia DR2}",
      journal = {Monthly Notices of the Royal Astronomical Society},
         year = 2019,
        month = apr,
       volume = {484},
       number = {3},
        pages = {3291-3306},
          doi = {10.1093/mnras/stz209},
archivePrefix = {arXiv},
       eprint = {1805.03653},
 primaryClass = {astro-ph.GA},
       adsurl = {https://ui.adsabs.harvard.edu/abs/2019MNRAS.484.3291T}
}

@ARTICLE{Trick_2022,
author = {{Trick}, Wilma H.},
title = "{Identifying resonances of the Galactic bar in Gaia DR2: II. Clues from angle space}",
journal = {Monthly Notices of the Royal Astronomical Society},
year = 2022,
month = jan,
volume = {509},
number = {1},
pages = {844-865},
doi = {10.1093/mnras/stab2866},
archivePrefix = {arXiv},
eprint = {2011.01233},
primaryClass = {astro-ph.GA},
adsurl = {https://ui.adsabs.harvard.edu/abs/2022MNRAS.509..844T}
}

@ARTICLE{Tremaine_Weinberg_1984,
       author = {{Tremaine}, S. and {Weinberg}, M.~D.},
        title = "{Dynamical friction in spherical systems.}",
      journal = {\mnras},
         year = 1984,
        month = aug,
       volume = {209},
        pages = {729-757},
          doi = {10.1093/mnras/209.4.729},
       adsurl = {https://ui.adsabs.harvard.edu/abs/1984MNRAS.209..729T}
}

@ARTICLE{agama,
author = {{Vasiliev}, Eugene},
title = "{AGAMA: action-based galaxy modelling architecture}",
journal = {Monthly Notices of the Royal Astronomical Society},
year = 2019,
month = jan,
volume = {482},
number = {2},
pages = {1525-1544},
doi = {10.1093/mnras/sty2672},
archivePrefix = {arXiv},
eprint = {1802.08239},
primaryClass = {astro-ph.GA},
adsurl = {https://ui.adsabs.harvard.edu/abs/2019MNRAS.482.1525V}
}

@ARTICLE{Vasiliev_2019,
       author = {{Vasiliev}, Eugene},
        title = "{Proper motions and dynamics of the Milky Way globular cluster system from Gaia DR2}",
      journal = {Monthly Notices of the Royal Astronomical Society},
         year = 2019,
        month = apr,
       volume = {484},
       number = {2},
        pages = {2832-2850},
          doi = {10.1093/mnras/stz171},
archivePrefix = {arXiv},
       eprint = {1807.09775},
 primaryClass = {astro-ph.GA},
       adsurl = {https://ui.adsabs.harvard.edu/abs/2019MNRAS.484.2832V}
}

@ARTICLE{Wang_etal2016,
author = {{Wang}, Y. and {Athanassoula}, E. and {Mao}, S.},
title = "{Orbital classification in an N-body bar}",
journal = {Monthly Notices of the Royal Astronomical Society},
archivePrefix = "arXiv",
eprint = {1609.02632},
year = 2016,
month = dec,
volume = 463,
pages = {3499-3512},
doi = {10.1093/mnras/stw2301}
}

@ARTICLE{Wozniak2020,
       author = {{Wozniak}, Herv{\'e}},
        title = "{Diffusion of radial action in a galactic disc}",
      journal = {\aap},
         year = 2020,
        month = oct,
       volume = {642},
          eid = {A207},
        pages = {A207},
          doi = {10.1051/0004-6361/202038959},
archivePrefix = {arXiv},
       eprint = {2009.01079},
 primaryClass = {astro-ph.GA},
       adsurl = {https://ui.adsabs.harvard.edu/abs/2020A&A...642A.207W}
}

@ARTICLE{Zozulia_etal2024a,
author = {{Zozulia}, Viktor D. and {Smirnov}, Anton A. and {Sotnikova}, Natalia Ya},
title = "{Positive Lynden-Bell derivative as a ticket to the bar trap?}",
journal = {Monthly Notices of the Royal Astronomical Society},
year = 2024,
month = mar,
doi = {10.1093/mnras/stae702},
archivePrefix = {arXiv},
eprint = {2403.08326},
primaryClass = {astro-ph.GA},
adsurl = {https://ui.adsabs.harvard.edu/abs/2024MNRAS.tmp..730Z}
}

@ARTICLE{Zozulia_etal2024b,
author = {{Zozulia}, Viktor D. and {Smirnov}, Anton A. and {Sotnikova}, Natalia Ya. and {Marchuk}, Alexander A.},
title = "{Boxy/peanut shaping of a mature galactic bar in action-angle space}",
journal = {\aap},
year = 2024,
month = dec,
volume = {692},
eid = {A145},
pages = {A145},
doi = {10.1051/0004-6361/202452088},
archivePrefix = {arXiv},
eprint = {2411.10391},
primaryClass = {astro-ph.GA},
adsurl = {https://ui.adsabs.harvard.edu/abs/2024A&A...692A.145Z}
}

@ARTICLE{Zozulia_etal2025,
       author = {{Zozulia}, Viktor D. and {Sotnikova}, Natalia Ya. and {Smirnov}, Anton A.},
        title = "{Phase-space distortion as a key to unraveling galactic bar buckling}",
      journal = {\aap},
         year = 2025,
        month = jun,
       volume = {698},
          eid = {L26},
        pages = {L26},
          doi = {10.1051/0004-6361/202554837},
archivePrefix = {arXiv},
       eprint = {2506.00631},
 primaryClass = {astro-ph.GA},
       adsurl = {https://ui.adsabs.harvard.edu/abs/2025A&A...698L..26Z}
}

@article{Cleveland1979Loess,
    author = {Cleveland, William S.},
    title = {Robust Locally Weighted Regression and Smoothing Scatterplots},
    journal = {Journal of the American Statistical Association},
    volume = {74},
    number = {368},
    pages = {829-836},
    year = {1979},
    doi = {10.1080/01621459.1979.10481038}
}

@article{SavitzkyGolay1964,
    author = {Savitzky, Abraham and Golay, Marcel J. E.},
    title = {Smoothing and Differentiation of Data by Simplified Least Squares Procedures},
    journal = {Analytical Chemistry},
    volume = {36},
    number = {8},
    pages = {1627-1639},
    year = {1964},
    doi = {10.1021/ac60214a047}
}

@article{HodrickPrescott1997,
    author = {Hodrick, Robert J. and Prescott, Edward C.},
    title = {Postwar {U.S.} Business Cycles: An Empirical Investigation},
    journal = {Journal of Money, Credit and Banking},
    volume = {29},
    number = {1},
    pages = {1-16},
    year = {1997},
    doi = {10.2307/2953682},
    url = {https://www.jstor.org/stable/2953682}
}

@ARTICLE{numpy,
       author = {{Harris}, Charles R. and {Millman}, K. Jarrod and {van der Walt}, St{\'e}fan J. and {Gommers}, Ralf and {Virtanen}, Pauli and {Cournapeau}, David and {Wieser}, Eric and {Taylor}, Julian and {Berg}, Sebastian and {Smith}, Nathaniel J. and {Kern}, Robert and {Picus}, Matti and {Hoyer}, Stephan and {van Kerkwijk}, Marten H. and {Brett}, Matthew and {Haldane}, Allan and {del R{\'\i}o}, Jaime Fern{\'a}ndez and {Wiebe}, Mark and {Peterson}, Pearu and {G{\'e}rard-Marchant}, Pierre and {Sheppard}, Kevin and {Reddy}, Tyler and {Weckesser}, Warren and {Abbasi}, Hameer and {Gohlke}, Christoph and {Oliphant}, Travis E.},
        title = "{Array programming with NumPy}",
      journal = {\nat},
         year = 2020,
        month = sep,
       volume = {585},
       number = {7825},
        pages = {357-362},
          doi = {10.1038/s41586-020-2649-2},
archivePrefix = {arXiv},
       eprint = {2006.10256},
 primaryClass = {cs.MS},
       adsurl = {https://ui.adsabs.harvard.edu/abs/2020Natur.585..357H}
}

@ARTICLE{scipy,
       author = {{Virtanen}, Pauli and {Gommers}, Ralf and {Oliphant}, Travis E. and {Haberland}, Matt and {Reddy}, Tyler and {Cournapeau}, David and {Burovski}, Evgeni and {Peterson}, Pearu and {Weckesser}, Warren and {Bright}, Jonathan and {van der Walt}, St{\'e}fan J. and {Brett}, Matthew and {Wilson}, Joshua and {Millman}, K. Jarrod and {Mayorov}, Nikolay and {Nelson}, Andrew R.~J. and {Jones}, Eric and {Kern}, Robert and {Larson}, Eric and {Carey}, C.~J. and {Polat}, {\.I}lhan and {Feng}, Yu and {Moore}, Eric W. and {VanderPlas}, Jake and {Laxalde}, Denis and {Perktold}, Josef and {Cimrman}, Robert and {Henriksen}, Ian and {Quintero}, E.~A. and {Harris}, Charles R. and {Archibald}, Anne M. and {Ribeiro}, Ant{\^o}nio H. and {Pedregosa}, Fabian and {van Mulbregt}, Paul and {SciPy 1.  0 Contributors}},
        title = "{SciPy 1.0: fundamental algorithms for scientific computing in Python}",
      journal = {Nature Medicine},
         year = 2020,
        month = feb,
       volume = {17},
        pages = {261-272},
          doi = {10.1038/s41592-019-0686-2},
archivePrefix = {arXiv},
       eprint = {1907.10121},
 primaryClass = {cs.MS},
       adsurl = {https://ui.adsabs.harvard.edu/abs/2020NatMe..17..261V}
}



\end{document}